\documentclass[aps,pre,reprint,amsmath,amssymb,superscriptaddress,longbibliography,floatfix]{revtex4-2}

\usepackage{bm}
\usepackage{mathtools}
\usepackage{graphicx}
\usepackage{xcolor}
\usepackage{microtype}
\usepackage{hyperref}
\hypersetup{colorlinks=true,citecolor=blue!45!black,linkcolor=blue!45!black,urlcolor=blue!45!black}

\allowdisplaybreaks
\newcommand{\Th}{T_{\mathrm h}}
\newcommand{\Tc}{T_{\mathrm c}}
\newcommand{\dd}{\,\mathrm d}
\newcommand{\PP}{\mathcal P}
\newcommand{\cA}{\mathcal A}
\newcommand{\cG}{\mathcal G}
\newcommand{\cR}{\mathcal R}
\newcommand{\ep}{\dot e_{p}}
\newcommand{\hd}{\dot h_{d}}
\newcommand{\st}{\mathrm{st}}
\newcommand{\sgn}{\operatorname{sgn}}
\newcommand{\ellref}{\ell_{\mathrm r}}

\begin{document}

\title{Thermodynamic optimization of thermal landscapes and energy barriers in a Brownian heat engine}

\author{Mesfin Asfaw Taye}
\email{tayem@wlac.edu}
\affiliation{West Los Angeles College, Science Division, 9000 Overland Avenue, Culver City, California 90230, USA}
\date{August 31, 2026}
\begin{abstract}
Spatial temperature fields in Brownian heat engines are commonly prescribed \emph{a priori}, and the resulting transport and thermodynamic properties are then calculated. Here we formulate the complementary inverse-design problem: determining the temperature profile and barrier height that optimize a chosen thermodynamic objective. We consider an overdamped Brownian particle in a symmetric triangular periodic potential under a constant opposing load and derive the exact stationary current and probability density for an arbitrary bounded temperature field, $\Tc\le T(x)\le\Th$. In the quasistatic limit, the efficiency becomes an exact functional of two inverse-temperature integrals over the uphill and downhill branches. Its rigorous global maximum under the pointwise temperature bounds is $\eta_{\max}=1-\Tc/\Th$, attained uniquely, up to sets of measure zero, by the hot-uphill/cold-downhill piecewise-constant profile. At finite current, however, the optimization changes qualitatively because the current is determined jointly by the cycle affinity and a nonlocal transport resistance. We derive the exact functional gradient and the corresponding box-constrained optimality conditions, showing that the current- or power-maximizing profile generally differs from the quasistatic efficiency optimum. For any prescribed temperature field, the current-maximizing barrier satisfies an exact balance between the marginal gain in thermal rectification and the marginal increase in transport resistance, with the characteristic estimate $U_0^*\simeq T_{\rm act}$, where $T_{\rm act}^{-1}=(2/L)\int_0^{L/2}\dd x/T(x)$. The steady-state entropy balance yields $\ep=\hd=J\cA$, whereas maximization of configurational Shannon entropy selects the potential-compensating profile $T_S(x)=T_0-U_s(x)$. Finally, the exponential temperature profile is identified as the unique fixed-endpoint minimizer of the squared logarithmic temperature gradient and as the smooth-profile limit of a regularized optimization. These results demonstrate that efficiency, current, power, dissipation, and configurational entropy generally select distinct optimal thermal landscapes.
\end{abstract}

\maketitle

\section{Introduction}
\label{sec:intro}

Brownian motors and Brownian heat engines are paradigmatic nonequilibrium systems that convert thermal fluctuations into directed motion and useful work \cite{Reimann2002,Hanggi2009,Sekimoto2010,Seifert2012,Taye2026BrownianMotorsBrownian}. They provide minimal yet physically informative models of energy conversion at microscopic scales, where thermal fluctuations cannot be neglected and may instead be used as a source of transport. When different regions of a reaction coordinate are maintained at different temperatures, local detailed balance is broken. A particle moving in a periodic potential can then acquire a nonzero average velocity and perform work against an opposing load, even when the potential itself is spatially symmetric \cite{Landauer1975,Buttiker1987,Derenyi1999,Hondou2000,Asfaw2002AdjustableBrownianHeat,Asfaw2005EnergeticsSimpleMicroscopic,Asfaw2007ExploringOperationTiny,Asfaw2008ModelingEfficientBrownian}. The spatial temperature field---specifically, where heat is supplied and where it is removed along the potential---is therefore not merely a boundary condition. It is the principal thermodynamic resource that determines the direction, strength, and energetic performance of the engine.

Previous studies have established the behavior of Brownian heat engines for several prescribed temperature arrangements. Piecewise-constant hot--cold profiles produced exactly solvable models and enabled the maximum-power analysis of Asfaw and Bekele \cite{AsfawBekele2004}, and related piecewise thermal arrangements were analyzed in Refs.~\cite{Asfaw2013EffectThermalInhomogeneity,Asfaw2014ThermodynamicFeatureBrownian,Taye2015ExactAnalyticalThermodynamic,Taye2015EffectTemperatureDependence,Duki2015EffectTemperatureViscous}. Brownian transport in a linearly varying temperature field was investigated in Refs.~\cite{Taye2017,Taye2021BrownianMotorsArranged}, including exact time-dependent solutions \cite{Taye2022ExactTimeDependent2,Taye2023TimeDependentSolutions}. More recently, an exponentially decreasing temperature profile was shown to yield the Curzon--Ahlborn form
\begin{equation}
\eta=1-\sqrt{\frac{\Tc}{\Th}}
\end{equation}
exactly in the quasistatic limit \cite{TayeExp2025,CurzonAhlborn1975}. This result established a direct connection between a continuous microscopic temperature field and a central efficiency expression of endoreversible thermodynamics. In a complementary variational approach, the periodic temperature field was held fixed while the potential landscape was optimized \cite{Berger2009}. Taken together, these studies demonstrate that thermal geometry strongly influences current, sustainable load, efficiency, power, and dissipation.

Despite this progress, the temperature field has generally been specified before the dynamics are solved. Such studies address the forward problem: given a profile $T(x)$, determine the current $J$, efficiency $\eta$, entropy production, and other thermodynamic properties. The corresponding inverse problem is fundamentally different: given a thermodynamic objective, determine the temperature profile $T^*(x)$ that optimizes it. Evaluating a small number of assumed profiles---such as piecewise-constant, linear, or exponential functions---cannot solve this problem rigorously because $T(x)$ belongs to an infinite-dimensional function space. A comparison among selected profiles can identify the best member of that limited set, but it cannot establish the global optimum over all admissible temperature fields.

The answer also depends on which temperature fields are physically or mathematically admissible. Under a pointwise temperature constraint, a discontinuous hot--cold piecewise-constant profile is permitted. In an experimental device, however, finite thermal conductivity, limited spatial resolution, and the cost of sustaining steep thermal gradients may require $T(x)$ to be continuous or smooth. Fixed endpoint temperatures, monotonicity, finite-gradient conditions, and thermal-control penalties can therefore change the optimal profile. It is essential to distinguish the ideal optimum obtained under minimal constraints from the realizable optimum obtained when spatial smoothness or finite-conductance restrictions are included.

Moreover, there is no single optimization problem because different observables reward different properties of the thermal landscape. A profile that maximizes quasistatic efficiency need not maximize finite current or output power. A profile that maximizes current need not maximize entropy production, and neither profile is generally expected to maximize the configurational Shannon entropy of the stationary probability density. Sharp thermal profiles can generate a large thermodynamic bias while simultaneously creating a kinetic bottleneck in a cold region. Smooth profiles may reduce that bottleneck and increase the current, even though they produce a smaller quasistatic efficiency. A systematic theory must therefore treat efficiency, current, power, dissipation, and probability spreading as distinct design objectives.

In the present work, both the temperature field $T(x)$ and the barrier height $U_0$ are treated as control variables for an overdamped Brownian particle moving in a symmetric triangular periodic potential under a constant opposing load $f$. The fundamental restriction on the temperature field is the pointwise box constraint
\begin{equation}
\Tc\le T(x)\le\Th ,
\label{eq:introbox}
\end{equation}
where $\Th>\Tc>0$. No piecewise-constant, linear, exponential, or other functional form is assumed at the outset. Fixed endpoint temperatures, monotonicity, and gradient penalties are introduced only when required by a particular optimization problem. This formulation converts the study of Brownian heat engines from a comparison of selected temperature profiles into an inverse-design problem over the complete admissible class.

The first principal result is an exact expression for the stationary current and probability density for an arbitrary bounded temperature field. The current separates naturally into a thermodynamic driving factor, determined by the cycle affinity, and a strictly positive nonlocal transport resistance. This decomposition makes the physical structure transparent: the affinity determines the direction and strength of the thermodynamic bias, whereas the resistance measures the kinetic difficulty of traversing the entire spatial period.

In the quasistatic limit, the stall force and efficiency reduce to exact functionals of two inverse-temperature integrals,
\begin{equation}
A[T]=\int_0^{L/2}\frac{\dd x}{T(x)},
\qquad
B[T]=\int_{L/2}^{L}\frac{\dd x}{T(x)},
\end{equation}
defined over the uphill and downhill branches, respectively. The efficiency is
\begin{equation}
\eta[T]=1-\frac{A[T]}{B[T]}.
\end{equation}
A direct global inequality and an independent functional-derivative calculation show that the unique box-constrained maximizer, apart from changes on sets of zero measure, is the hot-uphill/cold-downhill piecewise-constant profile. Its ideal configurational efficiency is
\begin{equation}
\eta_{\max}=1-\frac{\Tc}{\Th}.
\end{equation}
This result establishes the global optimum without restricting the calculation to any assumed family of trial profiles.

The exponential temperature field has a different but equally precise variational role. Although it does not maximize the box-constrained quasistatic efficiency, it is the unique fixed-endpoint minimizer of the squared logarithmic temperature gradient,
\begin{equation}
\int_0^L\left[\partial_x\ln T(x)\right]^2\dd x.
\end{equation}
It is therefore the smoothest profile when temperature variations are measured relative to the local temperature. This result explains why the exponential profile occupies a distinguished position among continuous thermal landscapes and clarifies its connection to the previously obtained Curzon--Ahlborn-type efficiency \cite{TayeExp2025}.

At finite current, the optimization changes qualitatively. A local variation of $T(x)$ changes not only the cycle affinity but also the transport resistance throughout the cell. The exact functional gradient of the current therefore contains a genuinely nonlocal term, and its sign cannot be determined solely from the local temperature or potential slope. Consequently, the finite-current or power-maximizing profile is not generally the hot--cold piecewise-constant profile that maximizes quasistatic efficiency. The present work derives the exact functional gradient and the corresponding box-constrained optimality conditions, thereby reducing finite-current thermal design to a well-defined nonlocal variational problem.

The barrier height introduces an additional design degree of freedom. For any prescribed temperature profile, the cycle affinity is exactly linear in $U_0$. The current-maximizing barrier satisfies an exact balance condition: the marginal increase in thermal rectification produced by raising the barrier must equal the associated marginal increase in transport resistance. A physically transparent activation estimate gives
\begin{equation}
U_0^*\simeq T_{\rm act},
\qquad
T_{\rm act}^{-1}
=\frac{2}{L}\int_0^{L/2}\frac{\dd x}{T(x)},
\end{equation}
which identifies the inverse-temperature weight of the uphill branch as the characteristic scale controlling the optimal barrier.

The thermodynamic and information-theoretic objectives are also shown to select different designs. A consistent steady-state entropy balance gives
\begin{equation}
\ep=\hd=J\cA,
\end{equation}
so entropy production and entropy extraction have the same optimizer, although that optimizer generally differs from the efficiency- and current-maximizing profiles. By contrast, maximum configurational Shannon entropy is obtained by flattening the stationary probability density. Whenever it is compatible with the temperature bounds, this condition selects the potential-compensating family
\begin{equation}
T_S(x)=T_0-U_s(x),
\end{equation}
which has a geometry fundamentally different from the hot-uphill/cold-downhill efficiency optimum. Finally, introducing a logarithmic-gradient penalty produces a regularized optimization that continuously connects the ideal piecewise-constant profile to the smooth exponential field as the cost of spatial temperature variation is increased.

Because the temperature varies in space, the thermal noise is multiplicative and the associated entropy bookkeeping requires particular care \cite{Polettini2013,Celani2012,BoCelani2013,Taye2016,Taye2020,Taye2021,Taye2015ExactAnalyticalExpressions,Taye2024ExactTimeDependent,taye2025EntropyProductionThermodynamic,Taye2025ThermodynamicIrreversibilityUnderdamped,Taye2026EntropyProductionMacroscopic}. All entropy-production and entropy-extraction rates in this work are derived from the same Fokker--Planck current used to describe the dynamics. The resulting identities therefore belong consistently to the configurational overdamped theory. Kinetic-energy transport between regions of different temperature and the entropic anomaly arising in the small-mass limit are treated separately and are not included in the ideal configurational efficiency.

The remainder of the paper is organized as follows. Section~\ref{sec:model} introduces the model, specifies the stochastic convention, and derives the exact stationary current and probability density for an arbitrary bounded temperature field. Section~\ref{sec:stall} obtains the stall force and the quasistatic efficiency functional, establishes its global optimum, and evaluates it for the piecewise-constant, linear, and exponential profiles. Section~\ref{sec:finite} develops the finite-current optimization with respect to $T(x)$, $U_0$, and $f$. Section~\ref{sec:entropy} derives the entropy balance and optimizes the dissipation, and Sec.~\ref{sec:shannon} solves the Shannon-entropy problem. Section~\ref{sec:reg} introduces the regularized smooth-profile optimization, identifies the variational role of the exponential field, and relates the results to the exponential-temperature engine. Section~\ref{sec:implications} discusses the design principles that emerge and the limitations of the overdamped description, and Sec.~\ref{sec:conclusions} summarizes the main conclusions. Technical derivations are provided in the appendixes.

\section{Model and stationary state}
\label{sec:model}

\subsection{Dynamics and thermal control}
\label{sec:dynamics}

We consider one spatial period $0\le x\le L$ of the symmetric triangular potential
\begin{equation}
U_s(x)=
\begin{cases}
\dfrac{2U_0}{L}\,x, & 0\le x\le L/2,\\[5pt]
2U_0\left(1-\dfrac{x}{L}\right), & L/2\le x\le L,
\end{cases}
\label{eq:Us}
\end{equation}
with barrier height $U_0>0$ and constant slopes
\begin{equation}
U_s'(x)=
\begin{cases}
+k, & 0<x<L/2,\\
-k, & L/2<x<L,
\end{cases}
\qquad k\equiv\frac{2U_0}{L}.
\label{eq:slopes}
\end{equation}
A constant opposing load $f>0$ is included through the tilted potential
\begin{equation}
U(x)=U_s(x)+fx,
\qquad
U'(x)=U_s'(x)+f,
\label{eq:Utotal}
\end{equation}
so that the deterministic force on the particle is $-U'(x)$. We set $k_B=1$ and retain the viscous friction coefficient $\gamma$.

\begin{figure}[t]
\includegraphics[width=\columnwidth]{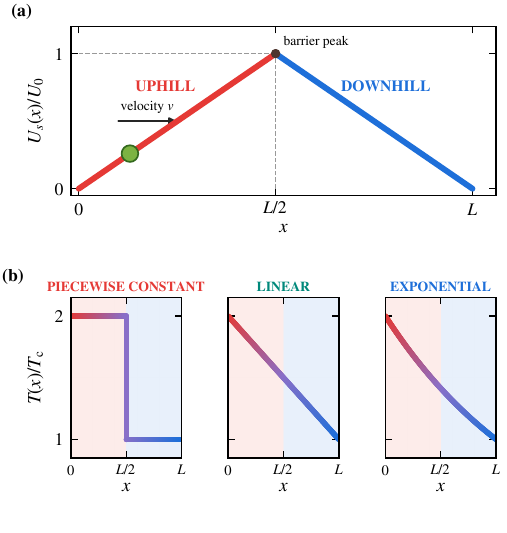}
\caption{(a) Symmetric triangular potential $U_s(x)$ of Eq.~\eqref{eq:Us}; the particle climbs the ascending (uphill) branch on $0<x<L/2$ and descends the downhill branch on $L/2<x<L$, and a net forward velocity $v=LJ$ arises when the thermal field breaks detailed balance. (b) The three canonical thermal landscapes obeying the box constraint of Eq.~\eqref{eq:box} for $\Th=2\Tc$: the piecewise-constant profile is hot on the entire uphill branch and cold on the entire downhill branch, while the linear and exponential profiles decrease continuously from $\Th$ at $x=0$ to $\Tc$ at $x=L$.}
\label{fig:model}
\end{figure}

The thermal control is a measurable positive field obeying the pointwise box constraint
\begin{equation}
0<\Tc\le T(x)\le\Th,
\qquad \Th>\Tc.
\label{eq:box}
\end{equation}
This broad class is used for the exact extremal results; fixed endpoints, monotone cooling, or gradient penalties are imposed only when stated. Figure~\ref{fig:model} shows the potential together with the three canonical profiles analyzed later.

To remain consistent with the exponential-profile engine of Ref.~\cite{TayeExp2025}, we adopt the divergence-form Smoluchowski equation
\begin{equation}
\frac{\partial P}{\partial t}
=\frac{1}{\gamma}\frac{\partial}{\partial x}
\left[U'(x)P+\frac{\partial}{\partial x}\bigl(T(x)P\bigr)\right]
=-\partial_xJ,
\label{eq:FPE}
\end{equation}
with probability current
\begin{equation}
J(x,t)=-\frac{1}{\gamma}
\left[U'(x)P+\partial_x\bigl(T(x)P\bigr)\right].
\label{eq:Jdef}
\end{equation}
At steady state $J$ is independent of $x$. Because the noise is multiplicative, Eq.~\eqref{eq:FPE} fixes one stochastic convention; the sensitivity of individual results to this choice is analyzed in Appendix~\ref{app:convention}.

A profile that decreases from $\Th$ at $x=0^+$ to $\Tc$ at $x=L^-$ is repeated periodically with a thermal reset at the cell boundary. For Eq.~\eqref{eq:FPE} the interface conditions across an ideal zero-width reset are continuity of the probability flux $J$ and continuity of the product $T(x)P(x)$; consequently, the density $P$ itself can jump inversely with $T$. Smooth periodic profiles are included as the special case without a reset. Closed-cycle line integrals below include the reset contribution whenever a logarithmic derivative of $T$ appears.

\subsection{Exact stationary current and density}
\label{sec:current}

At steady state, $\partial_tP=0$ forces $\partial_xJ=0$, so the current is a single unknown constant $J$, and Eq.~\eqref{eq:Jdef} becomes the ordinary differential equation
\begin{equation}
\frac{\dd}{\dd x}\bigl[T(x)P_s(x)\bigr]+U'(x)P_s(x)=-\gamma J,
\label{eq:steady0}
\end{equation}
where $P_s(x)$ is the stationary density. Equation~\eqref{eq:steady0} contains the unknown function in two places, inside the derivative as $TP_s$ and undifferentiated as $P_s$; the natural strategy is therefore to adopt the combination inside the derivative as the new dependent variable. Define
\begin{equation}
Q(x)=T(x)P_s(x),
\qquad
\Phi(x)=\int_0^x\frac{U'(z)}{T(z)}\dd z ,
\label{eq:QPhi}
\end{equation}
where $\Phi$, the thermal action, is the potential drop weighted at every point by the local inverse temperature; it will turn out to control both the sign of the current and the stall condition. Writing $P_s=Q/T$ in the second term of Eq.~\eqref{eq:steady0},
\begin{equation}
U'(x)P_s(x)=\frac{U'(x)}{T(x)}\,Q(x)=\Phi'(x)\,Q(x),
\label{eq:UPtoPhiQ}
\end{equation}
so the steady-state equation collapses to the standard first-order linear form
\begin{equation}
Q'(x)+\Phi'(x)\,Q(x)=-\gamma J .
\label{eq:Qode}
\end{equation}
Because the coefficient of $Q$ is exactly the derivative of $\Phi$, the integrating factor is $e^{\Phi(x)}$: multiplying Eq.~\eqref{eq:Qode} by it,
\begin{equation}
e^{\Phi}Q'+e^{\Phi}\Phi'Q
=\frac{\dd}{\dd x}\left[e^{\Phi(x)}Q(x)\right]
=-\gamma J\, e^{\Phi(x)},
\label{eq:IF}
\end{equation}
the left-hand side becomes a total derivative. Integrating Eq.~\eqref{eq:IF} from $0$ to $x$ and solving for $Q$,
\begin{equation}
Q(x)=e^{-\Phi(x)}
\left[Q(0)-\gamma J\int_0^xe^{\Phi(y)}\dd y\right].
\label{eq:Qsol0}
\end{equation}
Two constants remain undetermined, $Q(0)$ and $J$; they are fixed by the two conditions available, periodicity and normalization.

Denote
\begin{equation}
\Phi_L\equiv\Phi(L),
\qquad
I_\Phi\equiv\int_0^Le^{\Phi(y)}\dd y .
\label{eq:PhiLdef}
\end{equation}
The interface rule of Sec.~\ref{sec:model} states that the quantity continuous across the cell boundary is precisely $TP=Q$; periodicity therefore requires $Q(L^-)=Q(0^+)$. Evaluating Eq.~\eqref{eq:Qsol0} at $x=L$ and imposing this condition,
\begin{equation}
Q(0)=e^{-\Phi_L}\bigl[Q(0)-\gamma J\,I_\Phi\bigr],
\label{eq:Q0a}
\end{equation}
which, upon multiplying through by $e^{\Phi_L}$ and collecting the terms in $Q(0)$, gives $\bigl(e^{\Phi_L}-1\bigr)Q(0)=-\gamma J\,I_\Phi$ and hence
\begin{equation}
Q(0)=\frac{\gamma J\, I_\Phi}{1-e^{\Phi_L}} .
\label{eq:Q0}
\end{equation}
The remaining constant $J$ will be fixed below by normalization.
Substituting Eq.~\eqref{eq:Q0} back into Eq.~\eqref{eq:Qsol0} and using the identity
\begin{multline}
I_\Phi-\bigl(1-e^{\Phi_L}\bigr)\int_0^xe^{\Phi(y)}\dd y\\
=\int_x^Le^{\Phi(y)}\dd y
+e^{\Phi_L}\int_0^xe^{\Phi(y)}\dd y
\label{eq:periodicinner}
\end{multline}
gives the density in the compact form
\begin{equation}
P_s(x)=\frac{\gamma J}{1-e^{\Phi_L}}\,
\frac{e^{-\Phi(x)}}{T(x)}\,\cG(x),
\label{eq:PbeforeNorm}
\end{equation}
where the strictly positive kernel
\begin{equation}
\cG(x)=\int_x^Le^{\Phi(y)}\dd y
+e^{\Phi_L}\int_0^xe^{\Phi(y)}\dd y
\label{eq:Gkernel}
\end{equation}
collects the influence of the entire cell on the point $x$; both integrands in Eq.~\eqref{eq:Gkernel} are positive, so $\cG(x)>0$ everywhere. Define the strictly positive transport resistance as the integral of the density prefactor,
\begin{equation}
\cR[T,f]=\int_0^L\frac{e^{-\Phi(x)}}{T(x)}\,\cG(x)\dd x>0 .
\label{eq:R}
\end{equation}
Integrating Eq.~\eqref{eq:PbeforeNorm} over the cell then gives
\begin{equation}
\int_0^LP_s(x)\dd x
=\frac{\gamma J}{1-e^{\Phi_L}}\,\cR[T,f]=1,
\label{eq:normstep}
\end{equation}
and solving Eq.~\eqref{eq:normstep} for $J$ yields the exact stationary current and density for arbitrary bounded $T(x)$,
\begin{equation}
J[T,f]=\frac{1-e^{\Phi_L}}{\gamma\,\cR[T,f]},
\label{eq:Jexact}
\end{equation}
\begin{equation}
P_s[T](x)=\frac{e^{-\Phi(x)}\,\cG(x)}{T(x)\,\cR[T,f]} .
\label{eq:Pexact}
\end{equation}
No assumption of a piecewise-constant, linear, or exponential profile has entered these expressions; an alternative route to the same result is given in Appendix~\ref{app:current}.

Because $\cR>0$, the sign of the current is fixed entirely by the thermal action over one period,
\begin{equation}
\sgn J=-\sgn\Phi_L .
\label{eq:signJ}
\end{equation}
It is therefore natural to define the cycle affinity
\begin{equation}
\cA[T,f]\equiv-\Phi_L
=-\oint_0^L\frac{U'(x)}{T(x)}\dd x ,
\label{eq:affinity}
\end{equation}
in terms of which
\begin{equation}
J=\frac{1-e^{-\cA}}{\gamma\,\cR},
\qquad
J>0\;\Longleftrightarrow\;\cA>0 .
\label{eq:JAform}
\end{equation}
The heat-engine sector is $\cA>0$; at stall, $J=0$ and $\cA=0$.

Equation~\eqref{eq:JAform} exposes the structure that governs everything that follows. The numerator $1-e^{-\cA}$ measures the thermodynamic bias accumulated around one closed cycle: it vanishes when the weighted potential drop $\oint U'/T\,\dd x$ vanishes and saturates for strong driving. The denominator $\gamma\cR$ is a kinetic resistance accumulated over the complete cell; it depends nonlocally on $T(x)$ through both $\Phi$ and the kernel $\cG$. A local change of temperature can increase the affinity while simultaneously increasing the resistance, and it is exactly this competition that separates finite-current optimization from quasistatic efficiency optimization.

The conceptual content of this section can be summarized as follows. For an arbitrary bounded thermal field, the stationary state is fixed by a single scalar, the thermal action $\Phi_L$ accumulated around one period, together with one positive functional, the resistance $\cR$. The affinity $\cA=-\Phi_L$ decides whether the engine runs forward and how strongly it is driven; the resistance decides how fast probability can circulate for a given drive. Every design question addressed below is a question about how a proposed modification of $T(x)$ or $U_0$ redistributes weight between these two quantities. The quasistatic problem depends on the affinity alone, which is why it admits a closed-form global solution; the finite-current problems depend on both, which is why they become genuinely nonlocal.

\section{Quasistatic efficiency}
\label{sec:stall}

\subsection{Stall force and efficiency functional}
\label{sec:stallfunc}

Define the inverse-temperature integrals over the ascending and descending branches,
\begin{equation}
A[T]=\int_0^{L/2}\frac{\dd x}{T(x)},
\qquad
B[T]=\int_{L/2}^{L}\frac{\dd x}{T(x)} .
\label{eq:AB}
\end{equation}
Throughout, $A[T]$ and $B[T]$ denote these geometric functionals and are always distinct from the cycle affinity $\cA$ of Eq.~\eqref{eq:affinity}. Using the constant slopes of Eq.~\eqref{eq:slopes}, the thermal action over one period is
\begin{equation}
\Phi_L=(k+f)\,A+(f-k)\,B .
\label{eq:PhiAB}
\end{equation}
The stall force $f_\st$ is defined by $J=0$, i.e., by $\Phi_L=0$. Setting $f=f_\st$ in Eq.~\eqref{eq:PhiAB},
\begin{equation}
(k+f_{\st})A+(f_{\st}-k)B=0,
\label{eq:stall1}
\end{equation}
and collecting the terms proportional to $f_\st$ and to $k$,
\begin{equation}
f_{\st}\,(A+B)=k\,(B-A),
\label{eq:stall2}
\end{equation}
so that
\begin{equation}
f_{\st}[T]=k\,\frac{B[T]-A[T]}{A[T]+B[T]}
=\frac{2U_0}{L}\,\frac{B-A}{A+B} .
\label{eq:fstall}
\end{equation}
A positive stall force requires $B>A$: the engine can hold a load only if the downhill branch carries more inverse-temperature weight than the uphill branch.

For the energetics we use the standard idealized overdamped cycle bookkeeping of the Brownian-ratchet literature \cite{Derenyi1999,AsfawBekele2004,TayeExp2025}. In one net forward period at load $0<f<f_\st[T]$, the useful work and the configurational heat absorbed while climbing the barrier are
\begin{equation}
W(f)=fL,
\qquad
Q_{\rm in}(f)=U_0+\frac{fL}{2} .
\label{eq:WQfinite}
\end{equation}
The corresponding ideal cycle efficiency,
\begin{equation}
\eta_{\rm cyc}(f)=\frac{fL}{U_0+fL/2},
\qquad
\frac{\dd\eta_{\rm cyc}}{\dd f}
=\frac{LU_0}{(U_0+fL/2)^2}>0,
\label{eq:etaFiniteLoad}
\end{equation}
increases monotonically with $f$ at fixed $U_0$. In this bookkeeping the temperature profile does not enter at fixed load; its role is to determine how large a load the engine can sustain. The largest ideal efficiency of a given profile is therefore approached quasistatically, as $f\uparrow f_\st[T]$ with $J\to0^+$.

Setting $f=f_\st$ in Eq.~\eqref{eq:WQfinite} and inserting Eq.~\eqref{eq:fstall}, the work per period is
\begin{equation}
W_{\st}=f_\st L=2U_0\,\frac{B-A}{A+B},
\label{eq:WAB}
\end{equation}
while for the heat input the two contributions combine over a common denominator,
\begin{align}
Q_{{\rm in},\st}&=U_0+\frac{f_\st L}{2}
=U_0\left[1+\frac{B-A}{A+B}\right] \nonumber\\
&=U_0\,\frac{(A+B)+(B-A)}{A+B}
=\frac{2U_0B}{A+B} .
\label{eq:QAB}
\end{align}
Dividing Eq.~\eqref{eq:WAB} by Eq.~\eqref{eq:QAB}, both the barrier height $U_0$ and the common factor $(A+B)^{-1}$ cancel,
\begin{equation}
\eta[T]
=\frac{W_{\st}}{Q_{{\rm in},\st}}
=\frac{2U_0(B-A)/(A+B)}{2U_0B/(A+B)}
=\frac{B-A}{B},
\label{eq:etaDivision}
\end{equation}
and the quasistatic efficiency becomes the exact functional
\begin{equation}
\eta[T]
=1-\frac{A[T]}{B[T]}
=1-\frac{\displaystyle\int_0^{L/2}\dd x/T(x)}
{\displaystyle\int_{L/2}^{L}\dd x/T(x)} .
\label{eq:etaFunctional}
\end{equation}
The cancellation of $U_0$ is not accidental: both the sustainable load and the barrier climb are proportional to the same energy scale $U_0$, so their quasistatic ratio can depend only on the thermal geometry.

Equation~\eqref{eq:etaFunctional} is the first central result: the quasistatic efficiency depends on the entire thermal geometry only through the ratio of the inverse-temperature weights of the uphill and downhill branches. All details of where within each branch the temperature is high or low are irrelevant at stall; only the two integrals $A$ and $B$ survive. The finite-load expression \eqref{eq:etaFiniteLoad} is an ideal-cycle estimate; a full microscopic heat-current efficiency requires the underdamped kinetic-energy transfer discussed in Sec.~\ref{sec:limits}.

\subsection{Global optimum under the temperature bounds}
\label{sec:etaopt}

Equation~\eqref{eq:etaFunctional} already contains the solution of the efficiency design problem; this section makes it explicit, first by a direct global inequality and then by functional differentiation.

Because $A>0$ and $B>0$, maximizing $\eta=1-A/B$ is equivalent to minimizing the ratio $A/B$. The decisive observation is that the two integrals are independently controllable: $A$ involves the temperature only on the uphill half, and $B$ only on the downhill half. On the uphill branch the box constraint \eqref{eq:box} implies $1/\Th\le1/T(x)\le1/\Tc$, so $A$ is minimized pointwise by $T(x)=\Th$ for $0<x<L/2$, giving
\begin{equation}
A_{\min}=\frac{L}{2\Th}.
\label{eq:Amin}
\end{equation}
On the downhill branch, $B$ appears in the denominator of $A/B$ and should be maximized; the integrand $1/T$ is largest when $T$ is smallest, so $T(x)=\Tc$ for $L/2<x<L$, giving
\begin{equation}
B_{\max}=\frac{L}{2\Tc}.
\label{eq:Bmax}
\end{equation}
Hence, for every admissible profile,
\begin{equation}
\frac{A}{B}\ge\frac{A_{\min}}{B_{\max}}=\frac{\Tc}{\Th},
\qquad
\eta[T]\le1-\frac{\Tc}{\Th},
\label{eq:etaBound}
\end{equation}
with equality if and only if both pointwise choices are made simultaneously, up to sets of measure zero. Under the sole constraint $\Tc\le T(x)\le\Th$, the global maximizer is therefore the piecewise-constant profile
\begin{equation}
T_\eta^*(x)=
\begin{cases}
\Th, & 0<x<L/2,\\
\Tc, & L/2<x<L,
\end{cases}
\label{eq:TetaStar}
\end{equation}
with the ideal configurational efficiency
\begin{equation}
\eta_{\max}=1-\frac{\Tc}{\Th} .
\label{eq:CarnotForm}
\end{equation}
This is not merely a stationary point found by a derivative test; the inequality \eqref{eq:etaBound} establishes the global box-constrained optimum without assuming any trial family.

The same conclusion follows from a functional variation, which will also be needed for the finite-current problems below. A local variation means replacing $T(x)$ by $T_\epsilon(x)=T(x)+\epsilon\,\vartheta(x)$, where $\vartheta$ is an arbitrary admissible perturbation and $\epsilon\to0$; the functional derivative $\delta\eta/\delta T(x)$ is defined by $\delta\eta=\int_0^L[\delta\eta/\delta T(x)]\,\vartheta(x)\dd x$. From Eq.~\eqref{eq:AB},
\begin{align}
\delta A&=-\int_0^{L/2}\frac{\vartheta(x)}{T(x)^2}\dd x,
\label{eq:dA}\\
\delta B&=-\int_{L/2}^{L}\frac{\vartheta(x)}{T(x)^2}\dd x .
\label{eq:dB}
\end{align}
The signs follow from $\delta(1/T)=-\vartheta/T^2$: heating any point lowers the inverse-temperature integral through that point. Since $\eta=1-A/B$, the quotient rule gives
\begin{equation}
\delta\eta
=-\,\delta\!\left(\frac{A}{B}\right)
=-\frac{B\,\delta A-A\,\delta B}{B^2}
=-\frac{\delta A}{B}+\frac{A}{B^2}\,\delta B,
\label{eq:quotient}
\end{equation}
and inserting Eqs.~\eqref{eq:dA} and \eqref{eq:dB},
\begin{equation}
\delta\eta
=\frac{1}{B}\int_0^{L/2}\frac{\vartheta}{T^2}\dd x
-\frac{A}{B^2}\int_{L/2}^{L}\frac{\vartheta}{T^2}\dd x,
\label{eq:deltaEta}
\end{equation}
so that
\begin{equation}
\frac{\delta\eta}{\delta T(x)}=
\begin{cases}
\dfrac{1}{B\,T(x)^2}>0, & 0<x<L/2,\\[8pt]
-\dfrac{A}{B^2T(x)^2}<0, & L/2<x<L .
\end{cases}
\label{eq:gradEta}
\end{equation}
The gradient has a fixed sign on each half-cell: raising the local temperature anywhere on the uphill branch always increases the efficiency, and lowering it anywhere on the downhill branch always increases the efficiency. Because the derivative never vanishes in the interior of either branch, no smooth unconstrained stationary profile exists between the bounds, and the optimum is pushed onto the box boundary, reproducing Eq.~\eqref{eq:TetaStar}.

The physical content parallels the mathematics. The uphill half is where the particle must acquire energy to climb the barrier; placing the hottest permitted bath there maximizes thermal assistance where it is needed. The downhill half enters the stall ratio through $B$; placing the coldest permitted bath there maximizes the inverse-temperature asymmetry $B/A$ that sets the sustainable load. Every departure from $T=\Th$ on the uphill half or from $T=\Tc$ on the downhill half strictly lowers the ideal quasistatic efficiency on a set of nonzero measure.

Two remarks complete the picture. Without an engine-orientation constraint, the mathematical minimum of $\eta$ is the reversed piecewise-constant profile and is negative. If $T$ is instead required to decrease monotonically from hot to cold, then $A\le B$ and $\eta\ge0$, and the infimum $0$ is approached by an almost uniform profile whose endpoint changes are confined to narrow layers. Finally, the Carnot form \eqref{eq:CarnotForm} belongs to the ideal overdamped bookkeeping of Eqs.~\eqref{eq:WQfinite}--\eqref{eq:etaFiniteLoad}; it does not assert that a microscopic engine with kinetic heat leakage attains reversible Carnot performance (Sec.~\ref{sec:limits}).

\subsection{Piecewise-constant, linear, and exponential profiles}
\label{sec:profiles}

The three canonical profiles are now obtained as special cases of the single functional \eqref{eq:etaFunctional}, which makes their comparison exact rather than model dependent.

For the piecewise-constant profile \eqref{eq:TetaStar},
\begin{equation}
A_{\rm pc}=\frac{L}{2\Th},
\qquad
B_{\rm pc}=\frac{L}{2\Tc},
\label{eq:ABstep}
\end{equation}
so
\begin{equation}
\eta_{\rm pc}=1-\frac{\Tc}{\Th}.
\label{eq:etaStep}
\end{equation}

For the linear profile
\begin{equation}
T_{\rm lin}(x)=\Th-\frac{\Th-\Tc}{L}\,x,
\label{eq:Tlin}
\end{equation}
let $T_m=(\Th+\Tc)/2$ denote the midpoint temperature, $T_m=T_{\rm lin}(L/2)$. The integrand $1/T_{\rm lin}$ has the constant-slope antiderivative
\begin{equation}
\int\frac{\dd x}{T_{\rm lin}(x)}
=-\frac{L}{\Th-\Tc}\,\ln T_{\rm lin}(x)+{\rm const},
\label{eq:linAnti}
\end{equation}
so evaluating between the endpoints of each half-cell,
\begin{align}
A_{\rm lin}&=\frac{L}{\Th-\Tc}\,
\ln\frac{T_{\rm lin}(0)}{T_{\rm lin}(L/2)}
=\frac{L}{\Th-\Tc}\,\ln\!\left(\frac{\Th}{T_m}\right),
\label{eq:Alin}\\
B_{\rm lin}&=\frac{L}{\Th-\Tc}\,
\ln\frac{T_{\rm lin}(L/2)}{T_{\rm lin}(L)}
=\frac{L}{\Th-\Tc}\,\ln\!\left(\frac{T_m}{\Tc}\right),
\label{eq:Blin}
\end{align}
so
\begin{equation}
\eta_{\rm lin}=1-
\frac{\ln[2\Th/(\Th+\Tc)]}
{\ln[(\Th+\Tc)/(2\Tc)]} .
\label{eq:etaLin}
\end{equation}

\begin{figure}[t]
\includegraphics[width=\columnwidth]{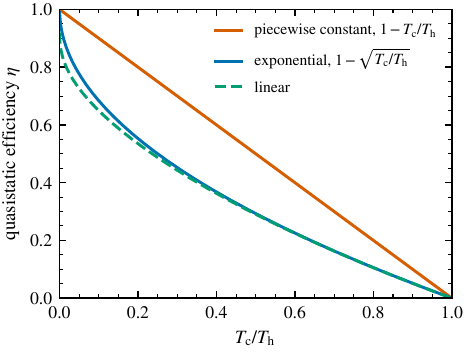}
\caption{Exact quasistatic efficiency, Eq.~\eqref{eq:etaFunctional}, versus $\Tc/\Th$ for the piecewise-constant [Eq.~\eqref{eq:etaStep}], exponential [Eq.~\eqref{eq:etaExp}], and linear [Eq.~\eqref{eq:etaLin}] profiles. The piecewise-constant profile attains the box-constrained maximum $1-\Tc/\Th$ for every temperature ratio; the exponential profile gives the Curzon--Ahlborn form.}
\label{fig:eta}
\end{figure}

For the exponentially decreasing profile studied in Ref.~\cite{TayeExp2025},
\begin{equation}
T_{\exp}(x)=\Th\exp\!\left[-\frac{x}{L}\ln\!\left(\frac{\Th}{\Tc}\right)\right],
\label{eq:Texp}
\end{equation}
write $r=\Th/\Tc$. The inverse profile is itself an exponential,
\begin{equation}
\frac{1}{T_{\exp}(x)}
=\frac{1}{\Th}\,e^{(x/L)\ln r},
\label{eq:invTexp}
\end{equation}
which integrates elementarily; over the two half-cells,
\begin{align}
A_{\exp}&=\frac{1}{\Th}\,\frac{L}{\ln r}
\Bigl[e^{(1/2)\ln r}-1\Bigr]
=\frac{L}{\Th\ln r}\bigl(\sqrt r-1\bigr),
\label{eq:Aexp}\\
B_{\exp}&=\frac{1}{\Th}\,\frac{L}{\ln r}
\Bigl[e^{\ln r}-e^{(1/2)\ln r}\Bigr]
=\frac{L}{\Th\ln r}\bigl(r-\sqrt r\bigr) .
\label{eq:Bexp}
\end{align}
Factoring $\sqrt r$ out of Eq.~\eqref{eq:Bexp} gives $r-\sqrt r=\sqrt r\,(\sqrt r-1)$, so the two integrals are locked into the exact ratio
\begin{equation}
B_{\exp}=\sqrt r\,A_{\exp}=\sqrt{\frac{\Th}{\Tc}}\,A_{\exp},
\label{eq:ABratioExp}
\end{equation}
and Eq.~\eqref{eq:etaFunctional} gives
\begin{equation}
\eta_{\exp}=1-\sqrt{\frac{\Tc}{\Th}},
\label{eq:etaExp}
\end{equation}
recovering exactly the Curzon--Ahlborn-type result of Ref.~\cite{TayeExp2025}. Figure~\ref{fig:eta} compares the three expressions. For every $0<\Tc/\Th<1$ the ordering is
\begin{equation}
\eta_{\rm pc}>\eta_{\exp}>\eta_{\rm lin},
\label{eq:etaOrdering}
\end{equation}
in accordance with the global bound \eqref{eq:etaBound}: any smooth monotone profile spends part of the uphill branch below $\Th$ and part of the downhill branch above $\Tc$, and both departures reduce the ratio $B/A$. The exponential profile lies above the linear one because its faster initial decrease keeps more of the low-temperature weight on the downhill branch.

Three conclusions emerge from this section. First, the quasistatic efficiency is controlled entirely by the geometric ratio $A/B$: the position of the hot and cold regions relative to the ascending and descending branches is what matters, and the fine structure of the temperature field within each branch is invisible at stall. Second, because $A$ and $B$ are controlled independently, the box-constrained optimum is not a compromise but a simultaneous saturation of both bounds, which necessarily produces a discontinuous piecewise-constant field. Third, the three canonical profiles are not competing models but three evaluations of one functional, and their ordering \eqref{eq:etaOrdering} follows from the global bound rather than from numerical comparison. The Carnot form obtained here is the maximum of the idealized configurational bookkeeping; whether it can be approached microscopically is a separate question addressed in Sec.~\ref{sec:limits}. What the quasistatic analysis cannot decide is how fast the engine runs, and this is the subject of the next section.

\section{Finite-current optimization}
\label{sec:finite}

At finite current the engine delivers the output power
\begin{equation}
\PP[T,f,U_0]=fL\,J[T,f,U_0] .
\label{eq:powerDef}
\end{equation}
At fixed load $f>0$, maximizing $\PP$ over either $T(x)$ or $U_0$ is equivalent to maximizing the current $J$. The decisive difference from the stall problem is visible in Eq.~\eqref{eq:JAform}: the current
\begin{equation}
J=\frac{1-e^{\Phi_L}}{\gamma\cR}
=\frac{1-e^{-\cA}}{\gamma\cR}
\label{eq:JdriveResistance}
\end{equation}
contains two independent ingredients, the cycle affinity $\cA$ measuring thermodynamic driving and the positive nonlocal resistance $\cR$ measuring kinetic opposition. A temperature change that increases $\cA$ can simultaneously increase $\cR$, so finite-current optimization cannot be decided from the affinity alone.

\subsection{Temperature variation}
\label{sec:Tvariation}

Write $N[T]\equiv1-e^{\Phi_L}$, so that $J=N/(\gamma\cR)$ and
\begin{equation}
\ln J=\ln N-\ln\gamma-\ln\cR .
\label{eq:lnJsplit}
\end{equation}
Working with $\ln J$ rather than $J$ is deliberate: the current is a ratio, and taking the logarithm converts the ratio into a difference, so the driving and resistance contributions to any derivative appear as separate, additively combined relative sensitivities. This bookkeeping will be used for every optimization below.

Consider first how the thermal action responds to a local change of temperature. Under $T\to T+\epsilon\vartheta$, the integrand of $\Phi$ varies through $\delta(1/T)=-\vartheta/T^2$, so to first order
\begin{equation}
\delta\Phi(x)
=\int_0^xU'(z)\,\delta\!\left(\frac{1}{T(z)}\right)\dd z
=-\int_0^x\frac{U'(z)}{T(z)^2}\,\vartheta(z)\dd z .
\label{eq:deltaPhiDeriv}
\end{equation}
A perturbation localized at $z$ contributes only if $z$ lies inside the integration range, i.e., only for $x>z$; reading off the kernel of Eq.~\eqref{eq:deltaPhiDeriv},
\begin{equation}
\frac{\delta\Phi(x)}{\delta T(z)}
=-\frac{U'(z)}{T(z)^2}\,\Theta(x-z),
\label{eq:deltaPhiT}
\end{equation}
where $\Theta$ is the Heaviside function: heating the point $z$ reduces the inverse-temperature weight $U'/T$ there and therefore shifts $\Phi(x)$ for every downstream point $x>z$. In particular, at the cell boundary,
\begin{equation}
\frac{\delta\Phi_L}{\delta T(z)}
=-\frac{U'(z)}{T(z)^2} .
\label{eq:deltaPhiL}
\end{equation}
The numerator therefore responds purely locally: by the chain rule, $\delta N=-e^{\Phi_L}\,\delta\Phi_L$, so
\begin{equation}
\frac{\delta\ln N}{\delta T(z)}
=\frac{1}{N}\frac{\delta N}{\delta T(z)}
=\frac{-e^{\Phi_L}}{1-e^{\Phi_L}}\,\frac{\delta\Phi_L}{\delta T(z)}
=\frac{e^{\Phi_L}}{1-e^{\Phi_L}}\,
\frac{U'(z)}{T(z)^2} .
\label{eq:deltaN}
\end{equation}

The resistance responds nonlocally, because the temperature enters Eq.~\eqref{eq:R} through three distinct routes: the explicit factor $1/T(x)$ in the integrand, the local weight $e^{-\Phi(x)}$, and the kernel $\cG(x)$, which itself contains $\Phi$ at every point of the cell. Varying the kernel \eqref{eq:Gkernel} term by term,
\begin{align}
\frac{\delta \cG(x)}{\delta T(z)}={}&
\int_x^L e^{\Phi(y)}
\frac{\delta\Phi(y)}{\delta T(z)}\dd y
+e^{\Phi_L}\frac{\delta\Phi_L}{\delta T(z)}
\int_0^x e^{\Phi(y)}\dd y \nonumber\\
&+e^{\Phi_L}\int_0^x e^{\Phi(y)}
\frac{\delta\Phi(y)}{\delta T(z)}\dd y,
\label{eq:deltaGT}
\end{align}
and varying Eq.~\eqref{eq:R}, including the explicit $1/T(x)$ factor at $x=z$,
\begin{align}
\frac{\delta\cR}{\delta T(z)}={}&
\int_0^L\frac{e^{-\Phi(x)}}{T(x)}
\left[-\cG(x)\frac{\delta\Phi(x)}{\delta T(z)}
+\frac{\delta \cG(x)}{\delta T(z)}\right]\dd x \nonumber\\
&-\frac{e^{-\Phi(z)}\cG(z)}{T(z)^2} .
\label{eq:deltaRT}
\end{align}
The three routes are now visible term by term: the first bracketed term is the variation of the weight $e^{-\Phi}$, the second is the variation of the kernel, and the final term outside the integral comes from the explicit $1/T$ factor evaluated at the varied point. Combining Eqs.~\eqref{eq:deltaN} and \eqref{eq:deltaRT} through Eq.~\eqref{eq:lnJsplit},
\begin{equation}
\frac{\delta\ln J}{\delta T(z)}
=\frac{e^{\Phi_L}}{1-e^{\Phi_L}}\,
\frac{U'(z)}{T(z)^2}
-\frac{1}{\cR}\,\frac{\delta\cR}{\delta T(z)},
\label{eq:dlnJdT}
\end{equation}
and hence
\begin{equation}
\frac{\delta J}{\delta T(z)}
=J\,\frac{\delta\ln J}{\delta T(z)},
\qquad
\frac{\delta\PP}{\delta T(z)}
=fL\,J\,\frac{\delta\ln J}{\delta T(z)} .
\label{eq:gradJ}
\end{equation}

The structure of Eq.~\eqref{eq:dlnJdT} should be read carefully. The first term is local: it depends only on the slope and temperature at the varied point. The second term is genuinely nonlocal: a change of $T$ at one point $z$ modifies the thermal action for all $x>z$, thereby the kernel $\cG(x)$ everywhere, thereby the stationary density, and thereby the resistance accumulated over the whole cell. Its sign depends on the complete temperature landscape, the barrier height, the load, and the stochastic convention (Appendix~\ref{app:convention}).

Under the pointwise bounds \eqref{eq:box}, a maximizing profile of any finite-current objective $\mathcal O[T]$ must satisfy the standard box-constrained (Karush--Kuhn--Tucker) conditions: wherever $\delta\mathcal O/\delta T(z)>0$ the optimal temperature saturates the upper bound, $T^*(z)=\Th$; wherever $\delta\mathcal O/\delta T(z)<0$ it saturates the lower bound, $T^*(z)=\Tc$; and an interior value $\Tc<T^*(z)<\Th$ is possible only where the functional derivative vanishes,
\begin{equation}
\frac{\delta\mathcal O}{\delta T(z)}=0
\qquad\text{on interior sets}.
\label{eq:KKTgeneric}
\end{equation}
For power, the relevant gradient is Eq.~\eqref{eq:gradJ}. The contrast with efficiency is fundamental: the efficiency gradient \eqref{eq:gradEta} has a fixed sign on each half-cell and yields a closed-form global optimum, whereas the sign of Eq.~\eqref{eq:dlnJdT} is not fixed locally. Consequently, no universal finite-barrier temperature profile maximizing current or power can be inferred from the bounds alone, and smooth profiles can carry more current than the sharp piecewise-constant profile over finite parameter ranges even though the piecewise-constant profile maximizes the quasistatic efficiency.

\subsection{Optimal barrier height}
\label{sec:Uopt}

For a fixed temperature profile, the affinity depends on the barrier height in a remarkably simple way. Inserting the slopes \eqref{eq:slopes} into Eq.~\eqref{eq:affinity} and using the half-cell integrals \eqref{eq:AB},
\begin{align}
\cA(U_0,f;T)
&=-(k+f)A+(k-f)B \nonumber\\
&=\frac{2U_0}{L}(B-A)-f(A+B) .
\label{eq:Abarrier}
\end{align}
Introducing the two positive profile functionals (for a hot-to-cold engine profile with $B>A$)
\begin{equation}
K_T\equiv\frac{2}{L}\,(B-A),
\qquad
M_T\equiv A+B,
\label{eq:KM}
\end{equation}
the affinity becomes exactly linear in the two control parameters,
\begin{equation}
\cA(U_0,f;T)=K_T\,U_0-M_T\,f .
\label{eq:AlinearU}
\end{equation}
The linearity is exact because the triangular slopes are constant on each half-cell, so $U_0$ and $f$ enter $U'/T$ only through the fixed weights $A$ and $B$. One immediate consequence is an onset threshold: for a positive load, forward engine transport requires $\cA>0$, i.e.,
\begin{equation}
U_0>U_{0,\mathrm{on}}(f;T)=\frac{M_T}{K_T}\,f
=\frac{Lf\,(A+B)}{2\,(B-A)} .
\label{eq:Uonset}
\end{equation}
Equation~\eqref{eq:Uonset} is the stall relation \eqref{eq:fstall} solved for $U_0$ rather than for $f$: below the threshold, the load drives the particle backward.

At fixed $T(x)$ and $f$, the barrier-optimization problem is one dimensional and can be solved exactly. Differentiating $\ln J$ rather than $J$ is again the efficient route: by Eq.~\eqref{eq:lnJsplit} the stationarity condition $\partial_{U_0}J=0$ becomes an equality of two separately computable relative sensitivities, one from the driving factor and one from the resistance. For the driving factor, $\ln N=\ln(1-e^{-\cA})$, and the linearity \eqref{eq:AlinearU} gives
\begin{equation}
\frac{\partial\cA}{\partial U_0}=K_T,
\label{eq:dAdU}
\end{equation}
so by the chain rule
\begin{equation}
\frac{\partial\ln N}{\partial U_0}
=\frac{e^{-\cA}}{1-e^{-\cA}}\,
\frac{\partial\cA}{\partial U_0}
=\frac{K_T}{e^{\cA}-1},
\label{eq:dlnNdU}
\end{equation}
where the last step multiplies numerator and denominator by $e^{\cA}$. Subtracting the resistance sensitivity,
\begin{equation}
\frac{\partial\ln J}{\partial U_0}
=\frac{K_T}{e^{\cA}-1}
-\frac{\partial\ln\cR}{\partial U_0} .
\label{eq:dlnJdU}
\end{equation}
An interior current maximum therefore satisfies the exact condition
\begin{equation}
\frac{K_T}{e^{K_TU_0^*-M_Tf}-1}
=\left.\frac{\partial\ln\cR}{\partial U_0}\right|_{U_0^*} .
\label{eq:UoptExact}
\end{equation}
This is the central finite-current result for the barrier: the left-hand side is the marginal gain in thermal rectification produced by increasing the barrier, and the right-hand side is the accompanying marginal increase in transport resistance. The optimum occurs when the marginal gain in thermal rectification produced by increasing the barrier is equal to the accompanying marginal increase in transport resistance; the fastest motor operates exactly at this balance.

The resistance derivative in Eq.~\eqref{eq:UoptExact} is itself fully explicit. Define
\begin{equation}
g(x)\equiv\frac{\partial U_s'(x)}{\partial U_0}
=\begin{cases}
2/L, & 0<x<L/2,\\
-2/L, & L/2<x<L,
\end{cases}
\label{eq:gdef}
\end{equation}
and the barrier sensitivity of the action,
\begin{equation}
\psi(x)\equiv\frac{\partial\Phi(x)}{\partial U_0}
=\int_0^x\frac{g(z)}{T(z)}\dd z .
\label{eq:psiU}
\end{equation}
Evaluating $\psi$ over the full period, the two half-cells contribute with opposite signs of $g$,
\begin{multline}
\psi(L)=\frac{2}{L}\left[
\int_0^{L/2}\frac{\dd z}{T(z)}
-\int_{L/2}^{L}\frac{\dd z}{T(z)}\right]\\
=\frac{2}{L}(A-B)=-K_T,
\label{eq:psiL}
\end{multline}
consistent with $\partial_{U_0}\Phi_L=-\partial_{U_0}\cA=-K_T$ from Eq.~\eqref{eq:AlinearU}.
Differentiating the kernel \eqref{eq:Gkernel} under the integral sign,
\begin{align}
\frac{\partial \cG(x)}{\partial U_0}={}&
\int_x^L\psi(y)\,e^{\Phi(y)}\dd y
+e^{\Phi_L}\psi(L)\int_0^xe^{\Phi(y)}\dd y \nonumber\\
&+e^{\Phi_L}\int_0^x\psi(y)\,e^{\Phi(y)}\dd y,
\label{eq:GU}
\end{align}
and therefore, from Eq.~\eqref{eq:R},
\begin{equation}
\frac{\partial\cR}{\partial U_0}
=\int_0^L\frac{e^{-\Phi(x)}}{T(x)}
\left[-\psi(x)\,\cG(x)+\frac{\partial \cG(x)}{\partial U_0}\right]\dd x .
\label{eq:RU}
\end{equation}
Thus Eq.~\eqref{eq:UoptExact} contains no unspecified quantity; for any proposed $T(x)$ it is a one-dimensional root problem in $U_0$, evaluated by quadratures (the general parameter derivative is given in Appendix~\ref{app:loadDerivative}). A local maximum additionally requires $\partial^2 J/\partial U_0^2<0$ at $U_0^*$, which is conveniently tested through
\begin{equation}
\frac{\partial^2\ln J}{\partial U_0^2}
=-\frac{K_T^2\,e^{\cA}}{(e^{\cA}-1)^2}
-\frac{\partial^2\ln\cR}{\partial U_0^2} .
\label{eq:curvatureU}
\end{equation}

The existence of an interior optimum is physically expected from the two limits. As $U_0\to0$ the periodic potential flattens, thermal rectification disappears, and $J\to0$; as $U_0\to\infty$ barrier crossing becomes strongly activated and again $J\to0$. By continuity, at zero load at least one intermediate maximum of $J(U_0)$ exists for any nontrivial hot-to-cold thermal arrangement. At $f>0$ the current is load-driven backward below the onset threshold \eqref{eq:Uonset}, becomes positive above it, and vanishes once more in the large-barrier trapping limit, so a positive engine-current maximum occurs above $U_{0,\mathrm{on}}$ whenever the engine sector is accessible.

\subsection{Activation estimate}
\label{sec:activation}

The exact root of Eq.~\eqref{eq:UoptExact} is the result to use for quantitative work; a transparent scale estimate follows by separating weak-barrier rectification from activated throughput. Define the effective inverse activation temperature of the uphill branch,
\begin{equation}
\beta_{\rm act}[T]
\equiv\frac{2}{L}\int_0^{L/2}\frac{\dd x}{T(x)}
=\frac{2A}{L},
\qquad
T_{\rm act}\equiv\beta_{\rm act}^{-1} .
\label{eq:betaAct}
\end{equation}
A minimal interpolation between the small-barrier linear growth of the rectification and the large-barrier Arrhenius suppression is
\begin{equation}
J(U_0,f)\simeq C(f,T)\,U_0
\exp\!\left[-\beta_{\rm act}
\left(U_0+\frac{fL}{2}\right)\right],
\label{eq:JactivationApprox}
\end{equation}
where the prefactor $C$ varies slowly compared with the exponential. The factor $U_0$ encodes the vanishing of rectification with the barrier, and the exponential encodes the Arrhenius cost of climbing $U_0+fL/2$ against the uphill inverse temperature $\beta_{\rm act}$. Taking the logarithm,
\begin{equation}
\ln J\simeq\ln C+\ln U_0-\beta_{\rm act}\left(U_0+\frac{fL}{2}\right),
\label{eq:lnJapprox}
\end{equation}
and differentiating term by term (the slowly varying $\ln C$ contributing negligibly),
\begin{equation}
\frac{\partial\ln J}{\partial U_0}
\simeq\frac1{U_0}-\beta_{\rm act},
\label{eq:dlnJapprox}
\end{equation}
which vanishes at a single point, so the estimated optimum is
\begin{equation}
U_0^*\simeq T_{\rm act} .
\label{eq:Uapprox}
\end{equation}
This is a useful scale estimate, not the exact optimal barrier: it does not reproduce the load shift contained in Eq.~\eqref{eq:UoptExact} and should be used as an initial physical estimate before the exact condition is solved. Its content is that the optimal barrier is set by the inverse-temperature weight of the uphill branch alone.

\begin{figure}[t]
\includegraphics[width=\columnwidth]{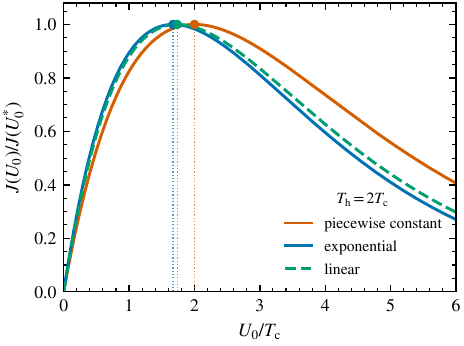}
\caption{Barrier-height dependence of the current in the activation estimate, Eq.~\eqref{eq:JactivationApprox}, for $\Th=2\Tc$ and zero load. Each curve is normalized by its own maximum, so the figure displays the shift of the optimal barrier with the temperature profile rather than an ordering of peak currents; markers and dotted verticals locate the optima $U_0^*=T_{\rm act}$ of Eq.~\eqref{eq:Uapprox}, which decrease from the piecewise-constant to the exponential to the linear profile [Eqs.~\eqref{eq:Ustep}--\eqref{eq:Uexp}]. Exact optima follow from Eq.~\eqref{eq:UoptExact}.}
\label{fig:barrier}
\end{figure}

For the three canonical profiles the activation scales follow directly from the half-cell integrals of Sec.~\ref{sec:profiles}. For the piecewise-constant profile, $\beta_{\rm act,pc}=1/\Th$ and
\begin{equation}
U_{0,\rm pc}^{*(\rm act)}=\Th .
\label{eq:Ustep}
\end{equation}
For the linear profile,
\begin{equation}
\beta_{\rm act,lin}=
\frac{2}{\Th-\Tc}\,\ln\frac{2\Th}{\Th+\Tc},
\quad
U_{0,\rm lin}^{*(\rm act)}=
\frac{\Th-\Tc}{2\ln\dfrac{2\Th}{\Th+\Tc}} .
\label{eq:Ulin}
\end{equation}
For the exponential profile, with $r=\Th/\Tc$,
\begin{equation}
\beta_{\rm act,exp}=
\frac{2(\sqrt r-1)}{\Th\ln r},
\qquad
U_{0,\rm exp}^{*(\rm act)}=
\frac{\Th\ln r}{2(\sqrt r-1)} .
\label{eq:Uexp}
\end{equation}
Because $\beta_{\rm act,pc}<\beta_{\rm act,exp}<\beta_{\rm act,lin}$, the piecewise-constant profile supports the largest optimal barrier and the linear profile the smallest; the entire temperature field thus changes the effective barrier scale that the particle experiences. Figure~\ref{fig:barrier} displays the normalized activation estimate for $\Th=2\Tc$; the exact quantitative optima follow from Eq.~\eqref{eq:UoptExact}, for which these scales serve as initial guesses, and the numerical procedure of Appendix~\ref{app:numerics} traces $U_0^*(f)$ for each profile without presupposing any current ordering.

\subsection{Power optimization}
\label{sec:poweropt}

At fixed nonzero load, $\PP=fLJ$ is proportional to the current, so the power-maximizing barrier coincides with the current-maximizing barrier,
\begin{equation}
U_{0,P}^*(f;T)=U_{0,J}^*(f;T) .
\label{eq:UPequalsUJ}
\end{equation}
If $U_0$ and $f$ are optimized jointly, Eq.~\eqref{eq:UoptExact} must be supplemented by the stationarity condition in the load. From Eq.~\eqref{eq:AlinearU},
\begin{equation}
\frac{\partial\cA}{\partial f}=-M_T,
\label{eq:dAdf}
\end{equation}
so, in complete analogy with Eq.~\eqref{eq:dlnJdU},
\begin{equation}
\frac{\partial\ln J}{\partial f}
=-\frac{M_T}{e^{\cA}-1}
-\frac{\partial\ln\cR}{\partial f} .
\label{eq:dlnJdf}
\end{equation}
Since $\partial_f\ln\PP=1/f+\partial_f\ln J$, the condition $\partial_f\PP=0$ reads
\begin{equation}
\frac1{f_P^*}
=\frac{M_T}{e^{\cA(U_0^*,f_P^*)}-1}
+\left.\frac{\partial\ln\cR}{\partial f}\right|_{U_0^*,f_P^*} .
\label{eq:fPexact}
\end{equation}
The joint optimum is determined by Eqs.~\eqref{eq:UoptExact} and \eqref{eq:fPexact} together; the required derivative $\partial_f\ln\cR$ is given by the same quadratures (Appendix~\ref{app:loadDerivative}). This pair makes the speed--power distinction precise: at fixed load, current and power select the same barrier, but once the load is also varied, power balances the gain in work per period against the loss of current.

Equation~\eqref{eq:JdriveResistance} explains why the profile with the largest stall force or quasistatic efficiency need not carry the largest finite current. The hot--cold piecewise-constant profile maximizes the efficiency because it minimizes $A/B$, but its extended cold region can create a strong kinetic bottleneck in $\cR$. Smooth linear or exponential fields distribute the cold region over the path and can reduce the dominant resistance. The exponentially decreasing-temperature engine of Ref.~\cite{TayeExp2025} indeed exhibits enhanced finite-current transport and entropy production over broad parameter ranges relative to sharper reference arrangements; this is fully compatible with the efficiency result of Sec.~\ref{sec:etaopt} because $\eta[T]$ and $J[T]$ are different functionals.

The ordering can also change with barrier height. In the strict Arrhenius limit the uphill exponential suppression is controlled by $\beta_{\rm act}$; since $\beta_{\rm act,pc}=1/\Th$ is the smallest of the three, the piecewise-constant-profile current is the least exponentially suppressed at asymptotically large $U_0$, even though a smooth profile can have a lower finite-barrier resistance and a larger current at moderate $U_0$. A crossing of current curves as $U_0$ grows is therefore physically natural rather than contradictory.

The finite-current results can be summarized in one statement: throughput is the ratio of a thermodynamic drive to a kinetic resistance, and every control variable---the temperature field, the barrier height, and the load---enters both. For the temperature field, this makes the optimization nonlocal and forbids any universal placement rule of the kind that governs the quasistatic efficiency. For the barrier height, it produces an interior optimum located by an exact marginal balance, Eq.~\eqref{eq:UoptExact}, whose scale is set by the uphill inverse temperature. For the load, it distinguishes maximum speed from maximum power. The practical consequence is that comparisons among thermal profiles are meaningful only when the barrier is either held fixed or optimized separately for each profile, and that a profile with the highest quasistatic efficiency may be a slower engine than a smooth profile over a broad range of barrier heights.

\section{Entropy production and extraction}
\label{sec:entropy}

\subsection{Entropy balance}
\label{sec:balance}

The configurational Shannon entropy relative to a fixed reference length $\ellref$ is
\begin{equation}
S(t)=-\int_0^LP(x,t)\ln[\ellref P(x,t)]\dd x ;
\label{eq:ShannonTime}
\end{equation}
the additive constant set by $\ellref$ affects no optimizing profile. Differentiating under the integral sign,
\begin{equation}
\dot S=-\int_0^L\bigl(\ln[\ellref P]+1\bigr)\,\partial_tP\dd x
=\int_0^L\bigl(\ln[\ellref P]+1\bigr)\,\partial_xJ\dd x,
\label{eq:SdotStart}
\end{equation}
where the continuity equation $\partial_tP=-\partial_xJ$ was used. The constant term integrates to $J(L)-J(0)=0$ around the closed cycle, and integrating the remaining term by parts,
\begin{equation}
\dot S=\Bigl[J\ln(\ellref P)\Bigr]_0^L
-\int_0^LJ\,\partial_x\ln P\dd x
=-\int_0^LJ\,\partial_x\ln P\dd x .
\label{eq:Sdot1}
\end{equation}
For a smooth periodic profile the boundary term vanishes by periodicity of $J$ and $P$. For a profile with an ideal thermal reset, the closed-cycle convention of Sec.~\ref{sec:model} assigns the reset jumps of both $\ln P$ and $\ln T$ to the corresponding line integrals; since $TP$ is continuous across the reset, the jump of $\ln P$ is exactly $-\Delta\ln T$, and the two jump contributions cancel in the combination appearing below, so no boundary term survives in either case.

The density gradient is now eliminated in favor of the current. Expanding the derivative in the current definition \eqref{eq:Jdef}, $\gamma J=-\bigl(U'P+T'P+TP'\bigr)$, and solving for $P'$,
\begin{equation}
TP'=-\gamma J-U'P-T'P ;
\label{eq:solveP}
\end{equation}
dividing Eq.~\eqref{eq:solveP} by $TP$,
\begin{equation}
\partial_x\ln P
=-\frac{\gamma J}{PT}-\frac{U'}{T}-\frac{T'}{T} .
\label{eq:lnPprime}
\end{equation}
Substituting Eq.~\eqref{eq:lnPprime} into Eq.~\eqref{eq:Sdot1}, the integrand becomes
\begin{equation}
-J\,\partial_x\ln P
=\frac{\gamma J^2}{PT}
+J\left(\frac{U'}{T}+\frac{T'}{T}\right),
\label{eq:integrandSplit}
\end{equation}
and integrating over the cell,
\begin{equation}
\dot S=
\gamma\int_0^L\frac{J^2}{PT}\dd x
+\int_0^LJ\left(\frac{U'}{T}+\frac{T'}{T}\right)\dd x .
\label{eq:balanceExpanded}
\end{equation}
The split in Eq.~\eqref{eq:integrandSplit} is not arbitrary: the first term is a positive-definite square generated by the current itself, while the second is linear in the current and changes sign with it; this is exactly the signature separating irreversible production from reversible exchange.
Consistently with the same current, we identify the entropy-production rate
\begin{equation}
\ep=\gamma\int_0^L\frac{J^2}{P_s(x)\,T(x)}\dd x\ \ge 0
\label{eq:epDef}
\end{equation}
and the entropy-extraction rate
\begin{equation}
\hd=-\int_0^LJ\left(\frac{U'}{T}+\frac{T'}{T}\right)\dd x,
\label{eq:hdDef}
\end{equation}
so that the balance takes the standard form
\begin{equation}
\dot S=\ep-\hd .
\label{eq:balance}
\end{equation}
Equation \eqref{eq:balance} has the usual flux-force structure: $\ep$ is a positive-definite quadratic form in the current, while $\hd$ is the entropy flow into the environment carried by the same current.

At a nonequilibrium steady state, $\dot S=0$ and $J$ is a constant. In Eq.~\eqref{eq:hdDef}, the term $\int_0^L(T'/T)\dd x$ is the closed-cycle integral of $\dd\ln T$ and vanishes for any periodic thermal cycle, including the reset jump of Sec.~\ref{sec:model}; the remaining term is $J\int_0^L(U'/T)\dd x=J\Phi_L$. Hence
\begin{equation}
\ep=\hd=-J\,\Phi_L=J\,\cA .
\label{eq:epJA}
\end{equation}
The same identity follows directly from the stationary equation, which provides an independent check of the bookkeeping. Dividing Eq.~\eqref{eq:Qode} by $Q=TP_s$,
\begin{equation}
\frac{Q'}{Q}=-\frac{\gamma J}{TP_s}-\Phi',
\label{eq:QoverQ}
\end{equation}
and integrating around the cell: the left-hand side is $\oint\dd\ln Q$, which vanishes because $Q=TP$ is the quantity that is continuous and periodic across the boundary, so
\begin{equation}
0=-\gamma J\int_0^L\frac{\dd x}{P_sT}-\Phi_L .
\label{eq:epAlt}
\end{equation}
Multiplying Eq.~\eqref{eq:epAlt} by $J$ and recognizing the first term as $-\ep$ from Eq.~\eqref{eq:epDef},
\begin{equation}
0=-\ep-J\Phi_L
\quad\Longrightarrow\quad
\ep=-J\Phi_L=J\cA,
\label{eq:epAlt2}
\end{equation}
reproducing Eq.~\eqref{eq:epJA}. Equation~\eqref{eq:epJA} states that the steady-state dissipation is the product of one flux, the cycle current $J$, and one conjugate force, the cycle affinity $\cA$.

At strict stall the configurational current vanishes and both rates in Eq.~\eqref{eq:epJA} vanish. This does not imply that an underdamped particle ceases to transport kinetic energy between hot and cold regions; that anomalous contribution lies outside Eq.~\eqref{eq:FPE} \cite{Celani2012,BoCelani2013} and is discussed in Sec.~\ref{sec:limits}.

\subsection{Optimization of dissipation}
\label{sec:epopt}

Equation~\eqref{eq:epJA} yields one exact structural statement immediately: at every nonequilibrium steady state of the adopted overdamped model,
\begin{equation}
\ep[T,f,U_0]=\hd[T,f,U_0],
\label{eq:epHdSame}
\end{equation}
so entropy production and entropy extraction are not independent optimization problems. Any variation of $T(x)$, $U_0$, or $f$ changes them identically, and their maximizers and minimizers coincide.

This equality does not imply that maximizing Shannon entropy minimizes either rate, nor that maximizing the affinity alone maximizes dissipation. Since $\ep=J\cA$, both the nonlocal transport factor $J$ and the local driving factor $\cA$ matter. The affinity gradient is purely local: from Eq.~\eqref{eq:affinity} and the variation used in Eq.~\eqref{eq:deltaPhiL},
\begin{equation}
\frac{\delta\cA}{\delta T(x)}=
\frac{U'(x)}{T(x)^2},
\label{eq:gradAffinity}
\end{equation}
whereas the current gradient is Eq.~\eqref{eq:gradJ}. By the product rule,
\begin{equation}
\frac{\delta\ep}{\delta T(x)}
=\cA\,\frac{\delta J}{\delta T(x)}
+J\,\frac{U'(x)}{T(x)^2},
\label{eq:gradEP}
\end{equation}
which at steady state is also the functional gradient of $\hd$. The box-constrained maximizing field obeys the optimality rule of Eq.~\eqref{eq:KKTgeneric} with the gradient \eqref{eq:gradEP}. Because the first term is nonlocal, there is no general finite-barrier statement forcing the $\ep$-maximizing field to coincide with the efficiency-maximizing piecewise-constant profile; maximum dissipation is a third, distinct design problem. We emphasize that maximizing entropy production is not in itself an engineering goal; identifying its extremizer serves to reveal how dissipation is distributed and how it relates to power; complementary scaling relations among current, activity, and entropy production are discussed in Refs.~\cite{Taye2025UnifiedNonequilibriumFramework,Taye2026UniversalThermodynamicInequality}.

For a prescribed $T(x)$ the barrier dependence is again transparent. Using $\cA=K_TU_0-M_Tf$ and Eq.~\eqref{eq:dlnJdU},
\begin{equation}
\frac{\partial\ln\ep}{\partial U_0}
=\frac{\partial\ln J}{\partial U_0}
+\frac{K_T}{\cA}
=\frac{K_T}{e^{\cA}-1}
-\frac{\partial\ln\cR}{\partial U_0}
+\frac{K_T}{\cA},
\label{eq:dlnEPdU}
\end{equation}
so an interior barrier maximizing both $\ep$ and $\hd$ satisfies
\begin{equation}
\frac{K_T}{e^{\cA}-1}
+\frac{K_T}{\cA}
=\frac{\partial\ln\cR}{\partial U_0} .
\label{eq:UoptEP}
\end{equation}
Comparing with the current condition \eqref{eq:UoptExact}, the additional positive term $K_T/\cA$ shows that the dissipation-optimal barrier is generically shifted toward larger thermodynamic driving: entropy production rewards affinity as well as throughput.

The minimum is simpler in the configurational overdamped model. Since $\ep\ge0$, any stall state with $J=0$ attains
\begin{equation}
\dot e_{p,\min}=\dot h_{d,\min}=0
\label{eq:epMinStall}
\end{equation}
within this description. As emphasized in Sec.~\ref{sec:limits}, an underdamped particle can retain kinetic-energy-mediated entropy production even when the configurational current vanishes, so Eq.~\eqref{eq:epMinStall} is not a claim of microscopic reversibility in the full phase-space model.

The entropy analysis adds a third design objective and simultaneously constrains it. The identity $\ep=\hd=J\cA$ shows that, within the configurational overdamped model, dissipation at steady state is completely specified by the same two quantities that specify the current, so maximizing entropy production is a finite-current problem of the same nonlocal type as maximizing power, with the additional weighting by the affinity that shifts the optimal barrier toward stronger driving. The balance therefore does not introduce a new independent optimizer; it locates dissipation within the affinity--resistance framework of Sec.~\ref{sec:finite}. What it does not determine is the shape of the stationary distribution itself, which is governed by a different functional and, as shown next, by a different optimal profile.

\section{Shannon entropy}
\label{sec:shannon}

\subsection{Uniform density and the compensating profile}
\label{sec:shannonuniform}

The stationary configurational entropy is the functional
\begin{equation}
S[T]=-\int_0^LP_s[T](x)\,\ln[\ellref P_s[T](x)]\dd x .
\label{eq:Sfunctional}
\end{equation}
For any normalized density on an interval of length $L$,
\begin{equation}
S\le\ln\frac{L}{\ellref},
\label{eq:Sbound}
\end{equation}
with equality if and only if the density is uniform,
\begin{equation}
P_s(x)=\frac1L ;
\label{eq:uniformP}
\end{equation}
a short relative-entropy proof is given in Appendix~\ref{app:Sbound}. The Shannon-entropy design problem is therefore solved by asking which admissible temperature fields render the stationary density uniform.

Substituting Eq.~\eqref{eq:uniformP} into the current definition \eqref{eq:Jdef}: with $P=1/L$ constant, the term $\partial_x(TP)$ reduces to $T'/L$, so
\begin{equation}
J=-\frac{1}{\gamma L}\left[U'(x)+T'(x)\right] .
\label{eq:Juniform}
\end{equation}
The left-hand side is a constant at steady state while the right-hand side is a function of position; consistency therefore requires
\begin{equation}
U'(x)+T'(x)=C
\label{eq:uniformCondition}
\end{equation}
for some constant $C$. The constant is fixed by integrating Eq.~\eqref{eq:uniformCondition} once around the ring for a continuous periodic thermal field. Using $U'=U_s'+f$ from Eq.~\eqref{eq:Utotal},
\begin{equation}
CL=\int_0^L\bigl(U_s'+f+T'\bigr)\dd x
=0+fL+0=fL,
\label{eq:Cdetermination}
\end{equation}
since $U_s$ and $T$ are periodic and their derivatives integrate to zero, while the load contributes $fL$. Hence $C=f$, and subtracting the load from Eq.~\eqref{eq:uniformCondition}, the spatial parts must cancel separately at every point,
\begin{equation}
T'(x)=-U_s'(x) .
\label{eq:Tprime}
\end{equation}
Integrating,
\begin{equation}
T_S^*(x)=T_0-U_s(x),
\label{eq:TSstar}
\end{equation}
where $T_0$ is an integration constant. This profile exactly compensates the spatial part of the deterministic drift; the resulting density is uniform, and
\begin{equation}
J_S=-\frac{f}{\gamma L},
\qquad
S_{\max}=\ln\frac{L}{\ellref} .
\label{eq:JSuniform}
\end{equation}
The temperature profile that maximizes configurational entropy therefore compensates the spatial potential rather than maximizing the thermodynamic bias responsible for directed transport. For a positive opposing load the unconstrained entropy maximum is not even a heat engine: the load drives the particle backward while the density is flattened.

\begin{figure}[!htbp]
\includegraphics[width=\columnwidth]{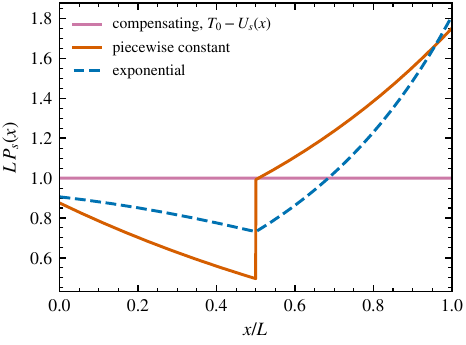}
\caption{Exact stationary densities $P_s(x)$ from Eq.~\eqref{eq:Pexact}, evaluated with the element-exact quadrature of Appendix~\ref{app:numerics}, for $L=\gamma=\Tc=1$, $\Th=2$, $U_0=0.8$, and $f=0.25$. The potential-compensating profile $T_0-U_s(x)$ with $T_0=1.9$ yields the exactly uniform density that attains the Shannon-entropy bound \eqref{eq:Sbound}; the piecewise-constant and exponential fields yield nonuniform densities and strictly smaller entropy. The piecewise-constant-profile density jumps at $x=L/2$ because the interface rule of Sec.~\ref{sec:model} makes $TP$, not $P$, continuous across a temperature discontinuity.}
\label{fig:Sdensity}
\end{figure}

For the triangular potential, Eq.~\eqref{eq:TSstar} reads explicitly
\begin{equation}
T_S^*(x)=
\begin{cases}
T_0-2U_0x/L, & 0<x<L/2,\\[4pt]
T_0-2U_0(1-x/L), & L/2<x<L,
\end{cases}
\label{eq:TStriangle}
\end{equation}
which is hottest at the potential minima and coldest at the barrier top---a geometry entirely different from the hot-uphill piecewise-constant profile of Sec.~\ref{sec:etaopt}. The profile obeys the box constraint \eqref{eq:box} provided $T_0-U_0\ge\Tc$ and $T_0\le\Th$ can hold simultaneously, i.e., provided
\begin{equation}
U_0\le\Th-\Tc .
\label{eq:Sfeasible}
\end{equation}
When Eq.~\eqref{eq:Sfeasible} holds, every member of the one-parameter family
\begin{equation}
T_S^*(x)=T_0-U_s(x),
\qquad
\Tc+U_0\le T_0\le\Th,
\label{eq:TSfamily}
\end{equation}
produces $P_s=1/L$ and attains the absolute bound $S_{\max}=\ln(L/\ellref)$. Conversely, within the periodic divergence-form model, any admissible profile attaining the bound must belong to this family up to changes on sets of measure zero: equality in Eq.~\eqref{eq:Sbound} forces $P_s=1/L$, Eqs.~\eqref{eq:Juniform}--\eqref{eq:uniformCondition} then force $T'=-U_s'$, and periodicity fixes the profile up to the constant $T_0$ constrained only by the bounds. Figure~\ref{fig:Sdensity} illustrates that the compensating family produces a flat stationary density, unlike the efficiency-optimal piecewise-constant profile or the exponential field.

\subsection{Constrained maximization}
\label{sec:shannonconstrained}

If $U_0>\Th-\Tc$, no bounded temperature can cancel the full potential, and the maximization becomes a genuine constrained problem. Its exact gradient can be obtained without differentiating the lengthy explicit density formula by an adjoint construction.

Let $\vartheta(x)=\delta T(x)$, $p(x)=\delta P_s(x)$, and $j=\delta J$ denote linked first-order variations. Linearizing the stationary equation \eqref{eq:steady0},
\begin{equation}
\frac{\dd}{\dd x}\bigl[T\,p+P_s\,\vartheta\bigr]+U'p=-\gamma j,
\qquad
\int_0^Lp\,\dd x=0,
\label{eq:linearizedP}
\end{equation}
where the second relation preserves normalization. The entropy variation is
\begin{multline}
\delta S=-\int_0^L\bigl(\ln[\ellref P_s(x)]+1\bigr)\,p(x)\dd x\\
=-\int_0^L\ln[\ellref P_s(x)]\,p(x)\dd x,
\label{eq:deltaS1}
\end{multline}
the constant term dropping by normalization. Equation~\eqref{eq:deltaS1} expresses $\delta S$ through the density response $p$, which depends on $\vartheta$ only implicitly through the linearized state equation \eqref{eq:linearizedP}. The adjoint method removes this implicit dependence by introducing a fixed dual field that trades the unknown response $p$ for the known perturbation $\vartheta$. Introduce a periodic adjoint field $\chi(x)$ satisfying
\begin{equation}
-T(x)\,\chi'(x)+U'(x)\,\chi(x)
=-\ln[\ellref P_s(x)]+\Lambda,
\label{eq:Sadjoint}
\end{equation}
with the gauge
\begin{equation}
\int_0^L\chi(x)\dd x=0 .
\label{eq:chiGauge}
\end{equation}
The left-hand side of Eq.~\eqref{eq:Sadjoint} is precisely the combination that will multiply $p$ after integration by parts, and the right-hand side is chosen to be the entropy weight of Eq.~\eqref{eq:deltaS1}; the constant $\Lambda$ enforces periodic solvability of the adjoint equation, and the gauge \eqref{eq:chiGauge} will eliminate the unknown current response $j$.

Multiply Eq.~\eqref{eq:linearizedP} by $\chi$ and integrate over the cell,
\begin{multline}
\int_0^L\chi\,\frac{\dd}{\dd x}\bigl[Tp+P_s\vartheta\bigr]\dd x
+\int_0^L U'\chi\,p\dd x\\
=-\gamma j\int_0^L\chi\dd x=0,
\label{eq:adjStep1}
\end{multline}
where the right-hand side vanishes by the gauge \eqref{eq:chiGauge}. Integrating the first term by parts, with the boundary term vanishing by periodicity,
\begin{equation}
\int_0^L\chi\,\frac{\dd}{\dd x}\bigl[Tp+P_s\vartheta\bigr]\dd x
=-\int_0^L\chi'\,\bigl[Tp+P_s\vartheta\bigr]\dd x,
\label{eq:adjStep2}
\end{equation}
so Eq.~\eqref{eq:adjStep1} rearranges to
\begin{equation}
\int_0^Lp\,\bigl[-T\chi'+U'\chi\bigr]\dd x
=\int_0^LP_s\,\chi'\,\vartheta\dd x .
\label{eq:adjStep3}
\end{equation}
Inserting the adjoint equation \eqref{eq:Sadjoint} on the left, the constant $\Lambda$ drops out by the normalization constraint $\int_0^Lp\,\dd x=0$, and what remains is exactly the entropy variation \eqref{eq:deltaS1},
\begin{equation}
\int_0^Lp\,\bigl(-\ln[\ellref P_s]+\Lambda\bigr)\dd x
=-\int_0^Lp\,\ln[\ellref P_s]\dd x=\delta S .
\label{eq:adjStep4}
\end{equation}
Combining Eqs.~\eqref{eq:adjStep3} and \eqref{eq:adjStep4},
\begin{equation}
\delta S=\int_0^LP_s(x)\,\chi'(x)\,\vartheta(x)\dd x,
\label{eq:deltaSAdj}
\end{equation}
so the exact functional gradient is
\begin{equation}
\frac{\delta S}{\delta T(x)}=P_s(x)\,\chi'(x) .
\label{eq:gradS}
\end{equation}
The box-constrained entropy maximizer therefore obeys the rule \eqref{eq:KKTgeneric} with this gradient: since $P_s>0$, interior (nonsaturated) temperature values require $\chi'=0$, and hot or cold saturated regions occur where the sign of $\chi'$ demands them.

A useful candidate consistent with these conditions is the clipped compensation profile
\begin{equation}
T_{S,\rm clip}(x;T_0)=
\min\bigl\{\Th,\max\bigl[\Tc,\;T_0-U_s(x)\bigr]\bigr\},
\label{eq:Tclip}
\end{equation}
with $T_0$ chosen to maximize $S$. It retains the exact compensating slope $T'=-U_s'$ wherever neither bound is active and saturates where full compensation is impossible. The numerical procedure of Appendix~\ref{app:numerics} verifies the gradient \eqref{eq:gradS} against finite differences and converges to this clipped structure for the triangular potential.

Finally, if the Shannon entropy is maximized subject to heat-engine operation, the inequality $\cA[T,f]\ge0$ must be adjoined. With a multiplier $\zeta\ge0$ and the gradients \eqref{eq:gradS} and \eqref{eq:gradAffinity}, stationarity of $S+\zeta\cA$ requires
\begin{equation}
\frac{\delta}{\delta T(x)}\bigl(S+\zeta\cA\bigr)
=P_s(x)\,\chi'(x)+\zeta\,\frac{U'(x)}{T(x)^2},
\qquad \zeta\cA=0,
\label{eq:SengineKKT}
\end{equation}
together with the box conditions. Unlike the unconstrained uniform-density result, the engine-constrained entropy maximizer is generally parameter dependent; Eq.~\eqref{eq:SengineKKT} with the adjoint equation \eqref{eq:Sadjoint} gives its complete characterization.

The Shannon-entropy problem has a qualitatively different character from all preceding objectives. Efficiency, current, power, and dissipation reward thermal asymmetry between the two branches, whereas configurational entropy rewards the cancellation of the potential by the temperature field, $T'=-U_s'$, which flattens the stationary density. The two designs are geometrically opposite: the efficiency optimum is hottest where the potential rises and coldest where it falls, while the entropy optimum is hottest at the potential minima and coldest at the barrier top. When the barrier exceeds the available temperature window, full compensation becomes impossible and the problem turns into a constrained adjoint problem whose solution saturates the bounds where compensation fails. A high-entropy stationary state is therefore neither a low-dissipation state nor a fast motor; it answers a different question.

\section{Smooth temperature profiles}
\label{sec:reg}

\subsection{Regularized optimization}
\label{sec:regopt}

The ideal piecewise-constant profile of Eq.~\eqref{eq:TetaStar} has an infinite gradient. It is mathematically admissible under the box constraint but experimentally idealized: laser heating, thermoplasmonic gradients, substrate heat diffusion, and interfacial thermal resistance all limit how sharply a temperature field can vary, and engineered thermal landscapes with localized hot or cold regions realize only finite gradients in practice \cite{Taye2026NoiseActivatedDopant}. A natural smooth-control problem penalizes sharp relative gradients,
\begin{equation}
\max_T\left\{\mathcal O[T]
-\mu\int_0^L\bigl(\partial_x\ln T\bigr)^2\dd x\right\},
\qquad\mu>0,
\label{eq:regularizedGeneric}
\end{equation}
where the objective $\mathcal O$ can be $\eta$, $\PP$, $\ep$, $\hd$, or $S$, and the penalty weight $\mu$ quantifies the physical cost assigned to steep logarithmic temperature gradients.

Let $y=\ln T$, so that $\delta T=T\,\delta y$ and the penalty is $\mu\int(y')^2\dd x$. Working in the variable $y$ is natural because the penalty is then quadratic and the box constraint remains a simple interval. Consider a variation $\delta y(x)$ vanishing at any fixed endpoints. The objective responds through the chain rule,
\begin{equation}
\delta\mathcal O
=\int_0^L\frac{\delta\mathcal O}{\delta T(x)}\,\delta T(x)\dd x
=\int_0^LT(x)\,\frac{\delta\mathcal O}{\delta T(x)}\,\delta y(x)\dd x,
\label{eq:regChain}
\end{equation}
while the penalty responds as
\begin{equation}
\delta\!\int_0^L(y')^2\dd x
=2\int_0^Ly'\,\delta y'\dd x
=-2\int_0^Ly''\,\delta y\dd x,
\label{eq:regPenaltyVar}
\end{equation}
after one integration by parts with vanishing boundary contributions. Stationarity of Eq.~\eqref{eq:regularizedGeneric} for arbitrary $\delta y$ therefore requires the pointwise condition
\begin{equation}
2\mu\,y''(x)+T(x)\,\frac{\delta\mathcal O}{\delta T(x)}=0 ,
\label{eq:universalRegularized}
\end{equation}
which holds in the interior of the admissible set for every choice of objective.
For the efficiency objective, inserting the gradient \eqref{eq:gradEta},
\begin{align}
2\mu\,y''+\frac{e^{-y}}{B}&=0,
\qquad 0<x<L/2,
\label{eq:regEta1}\\
2\mu\,y''-\frac{A}{B^2}\,e^{-y}&=0,
\qquad L/2<x<L,
\label{eq:regEta2}
\end{align}
where $A$ and $B$ are evaluated self-consistently on the solution. First integrals and matching conditions are derived in Appendix~\ref{app:regEta}.

\begin{figure}[!htbp]
\includegraphics[width=\columnwidth]{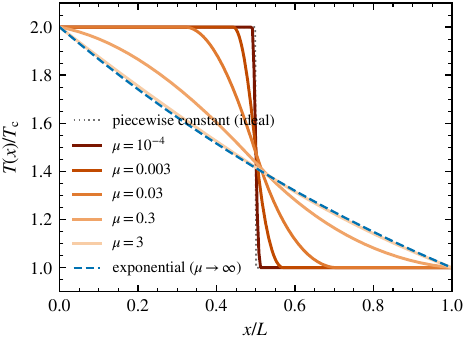}
\caption{Numerically optimized efficiency profiles from Eq.~\eqref{eq:regularizedGeneric} with fixed endpoints $T(0)=2\Tc$, $T(L)=\Tc$ and box bounds $\Tc\le T\le\Th$, for logarithmic-gradient penalties $\mu=10^{-4}$--$3$ (in units of $\Tc$; darker to lighter with increasing $\mu$), obtained by the gradient-based method of Appendix~\ref{app:numerics}. As $\mu\to0$ the optimum approaches the ideal hot/cold piecewise-constant profile (dotted); as $\mu$ grows it converges to the exponentially decreasing profile (dashed), the unique minimizer of the penalty functional at fixed endpoints.}
\label{fig:reg}
\end{figure}

\subsection{Sharp and smooth limits}
\label{sec:reglimits}

The two limits of the penalty weight connect the ideal and the smooth designs. As $\mu\to0^+$, the source terms in Eqs.~\eqref{eq:regEta1}--\eqref{eq:regEta2} dominate, the solution develops a narrow transition layer at $x=L/2$, and the profile approaches the piecewise-constant efficiency optimum \eqref{eq:TetaStar}.

The opposite limit selects the profile that minimizes the penalty alone, and this profile can be identified exactly. Define the logarithmic-gradient cost
\begin{equation}
C_{\log}[T]=\int_0^L\left(\frac{T'(x)}{T(x)}\right)^2\dd x
=\int_0^L[y'(x)]^2\dd x ,
\label{eq:Clog}
\end{equation}
with fixed endpoints $y(0)=\ln\Th$ and $y(L)=\ln\Tc$. The Euler--Lagrange equation of this quadratic functional is $y''=0$, whose fixed-endpoint solution is
\begin{equation}
y(x)=\ln\Th-\frac{x}{L}\ln\frac{\Th}{\Tc},
\label{eq:ylinear}
\end{equation}
which is exactly the exponential profile \eqref{eq:Texp}. The optimality is global rather than merely stationary: by the Cauchy--Schwarz inequality,
\begin{equation}
\int_0^L(y')^2\dd x
\ge\frac{1}{L}\left(\int_0^Ly'\dd x\right)^{\!2}
=\frac{1}{L}\ln^2\frac{\Th}{\Tc},
\label{eq:CSbound}
\end{equation}
with equality only for constant $y'$. As $\mu\to\infty$, Eqs.~\eqref{eq:regEta1}--\eqref{eq:regEta2} reduce to $y''\simeq0$, so the regularized optimum converges to this unique minimizer. By contrast, minimizing the absolute-gradient cost $\int_0^L(T')^2\dd x$ gives $T''=0$ and selects the linear profile \eqref{eq:Tlin}; the exponential is therefore the smoothest profile when temperature variations are measured relative to the local temperature, which is the measure relevant to a multiplicative-noise problem in which $1/T$ is the natural weight.

The three canonical profiles thus solve three different variational problems: the piecewise-constant profile maximizes the quasistatic efficiency $\eta[T]$; the exponential minimizes the relative-gradient cost \eqref{eq:Clog} at fixed endpoints; and the linear profile minimizes the absolute-gradient cost. The regularized problem interpolates continuously between the first two; Fig.~\ref{fig:reg} displays this interpolation. For regularized power, entropy production, or Shannon entropy, Eq.~\eqref{eq:universalRegularized} is used with the gradients \eqref{eq:gradJ}, \eqref{eq:gradEP}, or \eqref{eq:gradS}, respectively, so all five design problems share one state--adjoint boundary-value architecture.

\subsection{Relation to the exponential-temperature engine}
\label{sec:relation}

Reference~\cite{TayeExp2025} introduced the exponentially decreasing field \eqref{eq:Texp} and derived its exact Curzon--Ahlborn-type quasistatic efficiency. The present arbitrary-profile theory embeds that solvable model and clarifies its standing in four respects.

First, the functional \eqref{eq:etaFunctional} shows exactly why the exponential profile yields Eq.~\eqref{eq:etaExp}: its half-cell integrals obey $B_{\exp}=\sqrt{\Th/\Tc}\,A_{\exp}$. The same functional proves that the exponential is not the unconstrained efficiency maximizer; the hot--cold piecewise-constant profile attains the larger Carnot form.

Second, finite-current transport is governed by the affinity--resistance ratio \eqref{eq:JdriveResistance}, not by the efficiency alone. The earlier exponential engine found larger current and entropy production over broad finite-parameter regions than sharper reference arrangements \cite{TayeExp2025}; the present theory explains how this can occur, since a smooth distributed decrease of temperature can reduce the nonlocal resistance $\cR$ enough to compensate for a smaller affinity. The exact gradient \eqref{eq:dlnJdT} replaces any universal finite-barrier ranking by a parameter-dependent control problem.

Third, the barrier is an additional design coordinate. Equations~\eqref{eq:UoptExact} and \eqref{eq:UoptEP} show that each profile possesses its own current-, power-, and dissipation-optimal barrier. Profiles should therefore be compared either at one specified $U_0$ or after optimizing $U_0$ separately for each; mixing the two comparisons can obscure the actual transport advantage.

Fourth, the exponential profile has an independent variational meaning as the unique fixed-endpoint minimizer of the squared logarithmic gradient (Sec.~\ref{sec:reglimits}) and as the large-penalty limit of the regularized problem (Sec.~\ref{sec:reg}). The earlier exponential engine and the present theory thus answer complementary questions: the piecewise-constant profile is the ideal efficiency extremizer, the exponential is the minimum-relative-gradient field, and finite-current throughput is decided by the competition between affinity and resistance.

Taken together, the regularized problem and its two limits clarify the standing of the exponential field within the complete design picture. The piecewise-constant profile is the ideal quasistatic optimum under pointwise bounds alone; the exponential profile is the smoothest field in the relative sense of Eq.~\eqref{eq:Clog} and the natural end point when sharp thermal gradients carry a cost; and the finite-current advantage of smooth profiles reported earlier follows from the resistance term in the current, not from the efficiency. The single stationarity equation \eqref{eq:universalRegularized}, used with the appropriate gradient, places all five objectives within one state--adjoint framework, so the same numerical machinery serves the efficiency, power, dissipation, and entropy problems once the cost of thermal control is specified.

\section{Discussion}
\label{sec:implications}

\subsection{Design principles}
\label{sec:principles}

The results separate several distinct design mechanisms. Quasistatic efficiency depends only on the inverse-temperature geometry through $A/B$. Making the entire uphill branch as hot as possible and the entire downhill branch as cold as possible simultaneously minimizes $A$ and maximizes $B$, which is why the efficiency optimum is a sharp piecewise-constant profile.

Finite-current speed and power pose an additional kinetic question: how rapidly can probability traverse the complete cell? The affinity $\cA$ measures the thermodynamic bias around the cycle, while $\cR$ measures the time-consuming resistance generated by diffusion and barrier crossing. A sharp cold domain can raise the affinity yet create a bottleneck; a smooth field can relieve the bottleneck yet sacrifice some affinity. The finite-current optimum is a global balance, not a local placement rule.

Barrier height exposes the same competition in a single scalar variable. A very small barrier provides little thermal rectification, whereas a very large barrier traps the particle by Arrhenius activation. Equation~\eqref{eq:UoptExact} locates the intermediate optimum by equating the marginal rectification gain with the marginal resistance cost, and the approximate rule $U_0^*\simeq T_{\rm act}$ supplies the experimental starting scale before the exact root is computed.

Shannon entropy asks a third question: how uniform can the stationary density be made? Its maximizing field locally compensates the potential slope, $T'=-U_s'$, rather than maximizing directed transport. A high-entropy stationary state therefore need not be a low-dissipation state and need not be a fast motor.

Real thermal fields have finite spatial resolution. The regularized functional \eqref{eq:regularizedGeneric} allows implementation costs of sharp gradients to be included quantitatively without confusing them with the ideal box-constrained efficiency result, and it identifies the exponential profile as the natural smooth end point of the design family.

\subsection{Limitations of the overdamped description}
\label{sec:limits}

Three limitations of the present overdamped treatment must remain explicit.

First, the efficiency calculation uses the conventional configurational heat input of Eq.~\eqref{eq:WQfinite} and neglects the kinetic energy exchanged when the particle crosses between regions at different temperatures. Such a heat leak prevents true reversible Carnot operation in microscopic thermal ratchets \cite{Hondou2000,Celani2012,BoCelani2013}. Equation~\eqref{eq:CarnotForm} is therefore the maximum of the stated idealized configurational functional, not a claim that all microscopic heat currents vanish at the optimum.

Second, multiplicative thermal noise admits inequivalent stochastic representations. The present work fixes the divergence-form current \eqref{eq:Jdef} and derives $\ep$ and $\hd$ from that same equation; combining the current of one convention with the entropy-flow term of another would invalidate the balance \eqref{eq:balance}. Appendix~\ref{app:convention} shows which results are convention independent within a one-parameter family of conventions---the stall condition and the quasistatic efficiency functional are---and which are not, notably the finite-current resistance and all optimality conditions that involve it.

Third, the underdamped phase-space problem contains velocity degrees of freedom. A particle can carry kinetic energy from hot to cold even when the configurational current vanishes, producing the entropic anomaly of the small-mass limit \cite{Celani2012,BoCelani2013}. Optimizing the full underdamped entropy production is a separate control problem and need not select the same profile as the overdamped conditions of Sec.~\ref{sec:epopt}.

\section{Summary and conclusions}
\label{sec:conclusions}

The spatial temperature field and the barrier height of a Brownian heat engine have been treated as thermodynamic design variables. For an arbitrary bounded profile the stationary current and density are exact, Eqs.~\eqref{eq:Jexact} and \eqref{eq:Pexact}, and the quasistatic efficiency admits the simple exact representation $\eta[T]=1-A[T]/B[T]$ of Eq.~\eqref{eq:etaFunctional}. Its global box-constrained maximizer is the hot-uphill/cold-downhill piecewise-constant profile, with the ideal configurational value $1-\Tc/\Th$; the exponential profile plays a different exact role, giving the Curzon--Ahlborn form and uniquely minimizing the integrated squared logarithmic temperature gradient.

Finite-current performance obeys no such local rule. The current is a ratio of a driving numerator and a nonlocal resistance, its complete functional gradient \eqref{eq:dlnJdT} contains a term whose sign depends on the entire landscape, and consequently no universal finite-barrier current or power maximizer follows from the temperature bounds alone. This reconciles the exact efficiency result with the finite-current advantages of smooth distributed profiles, including the exponential field, reported previously \cite{TayeExp2025}. For any prescribed profile the barrier admits an exact optimization: the affinity is linear in $U_0$, $\cA=K_TU_0-M_Tf$, and the current-maximizing barrier satisfies the balance condition \eqref{eq:UoptExact} in which the marginal rectification gain equals the marginal resistance cost, with the activation scale $U_0^*\simeq T_{\rm act}$ as a physical estimate and Eq.~\eqref{eq:fPexact} completing the joint power optimum.

A consistent entropy balance gives $\ep=\hd=J\cA$ at steady state, so entropy production and entropy extraction always share one optimizer; that optimizer is itself a nonlocal finite-current problem, and the dissipation-optimal barrier \eqref{eq:UoptEP} generically differs from the maximum-speed barrier. Shannon entropy defines yet another design problem: whenever $U_0\le\Th-\Tc$, the potential-compensating family $T_0-U_s(x)$ renders the stationary density uniform and attains the absolute entropy bound, while stronger barriers lead to a constrained adjoint problem with clipped compensation.

The central physical conclusion is that no single spatial thermal profile is optimal for every observable. Efficiency, throughput, dissipation, and configurational entropy reward different combinations of thermal bias, kinetic accessibility, and probability spreading, and the arbitrary-profile efficiency functional, the exact finite-current optimality conditions, and the barrier balance law together provide a systematic framework for matching the thermal design to the quantity actually being optimized. Natural extensions include experimental thermal design with realistic gradient constraints; the full underdamped phase-space optimization, in which kinetic heat leaks modify both the attainable efficiency and the entropy bookkeeping; coupled degrees of freedom such as dimers and short polymer chains transported through inhomogeneous thermal landscapes \cite{Asfaw2010StochasticResonanceFlexible,Asfaw2012ExploringDynamicsDimer,Asfaw2014ThermallyActivatedBarrier,Taye2015RectifiedMotionShort,Taye2023TransportingShortPolymer,Taye2025DirectedTransportShort}; and hybrid engines that alternate between active and passive operation \cite{Taye2025CompetingActivePassive,Taye2025EntropyProductionThermodynamic,Taye2026ExactThermodynamicAnalysis}.

\appendix

\section{Alternative derivation of the exact current}
\label{app:current}

Starting from Eq.~\eqref{eq:Qsol0}, impose the interface condition $Q(L)=Q(0)$ directly:
\begin{align}
Q(0)&=e^{-\Phi_L}\bigl[Q(0)-\gamma J\,I_\Phi\bigr],
\label{eq:appQ1}\\
Q(0)\bigl(e^{\Phi_L}-1\bigr)&=-\gamma J\,I_\Phi,
\label{eq:appQ2}\\
Q(0)&=\frac{\gamma J\,I_\Phi}{1-e^{\Phi_L}} .
\label{eq:appQ3}
\end{align}
Substituting Eq.~\eqref{eq:appQ3} into Eq.~\eqref{eq:Qsol0},
\begin{align}
Q(x)&=\frac{\gamma J\,e^{-\Phi(x)}}{1-e^{\Phi_L}}
\left[I_\Phi-\bigl(1-e^{\Phi_L}\bigr)\int_0^xe^{\Phi(y)}\dd y\right]
\label{eq:appQ4}\\
&=\frac{\gamma J\,e^{-\Phi(x)}}{1-e^{\Phi_L}}\,\cG(x),
\label{eq:appQ5}
\end{align}
by the identity \eqref{eq:periodicinner}. Division by $T$ gives Eq.~\eqref{eq:PbeforeNorm}, and normalization gives
\begin{equation}
1=\frac{\gamma J}{1-e^{\Phi_L}}\,\cR,
\label{eq:appNorm}
\end{equation}
which is Eq.~\eqref{eq:Jexact}.

At stall, $\Phi_L\to0$, the kernel reduces to $\cG(x)\to I_\Phi$, and the density becomes the Boltzmann-like form
\begin{equation}
P_{\st}(x)=\frac{e^{-\Phi(x)}}{T(x)\,Z},
\qquad
Z=\int_0^L\frac{e^{-\Phi(x)}}{T(x)}\dd x .
\label{eq:Pstall}
\end{equation}

\section{Parameter derivatives of the transport resistance}
\label{app:loadDerivative}

For any parameter $p$ that enters the potential slope but not the prescribed temperature profile, define
\begin{equation}
\psi_p(x)=\frac{\partial\Phi(x)}{\partial p}
=\int_0^x\frac{\partial_pU'(z)}{T(z)}\dd z,
\qquad
\psi_{p,L}=\psi_p(L) .
\label{eq:psip}
\end{equation}
Differentiating the kernel \eqref{eq:Gkernel},
\begin{align}
\partial_p\cG(x)={}&\int_x^L\psi_p(y)\,e^{\Phi(y)}\dd y
+e^{\Phi_L}\psi_{p,L}\int_0^xe^{\Phi(y)}\dd y \nonumber\\
&+e^{\Phi_L}\int_0^x\psi_p(y)\,e^{\Phi(y)}\dd y,
\label{eq:appGp}
\end{align}
and therefore
\begin{equation}
\partial_p\cR
=\int_0^L\frac{e^{-\Phi(x)}}{T(x)}
\left[-\psi_p(x)\,\cG(x)+\partial_p\cG(x)\right]\dd x .
\label{eq:appRp}
\end{equation}
For $p=U_0$, $\partial_{U_0}U'=g(x)$ of Eq.~\eqref{eq:gdef} and $\psi_{U_0,L}=-K_T$, recovering Eqs.~\eqref{eq:GU}--\eqref{eq:RU}. For $p=f$, $\partial_fU'=1$, so
\begin{equation}
\psi_f(x)=\int_0^x\frac{\dd z}{T(z)},
\qquad
\psi_{f,L}=M_T .
\label{eq:psif}
\end{equation}
These quadratures supply $\partial_{U_0}\ln\cR$ and $\partial_f\ln\cR$ as required by the exact conditions \eqref{eq:UoptExact} and \eqref{eq:fPexact}; higher derivatives follow by one further differentiation of the same expressions.

\section{Relative-entropy proof of the Shannon bound}
\label{app:Sbound}

Let $P_u=1/L$ denote the uniform density. The Kullback--Leibler divergence of any normalized density $P$ from $P_u$ is nonnegative,
\begin{equation}
D(P\Vert P_u)=\int_0^LP(x)\,
\ln\frac{P(x)}{1/L}\dd x\ \ge0 .
\label{eq:KL}
\end{equation}
Expanding the logarithm,
\begin{equation}
D(P\Vert P_u)=\int_0^LP\ln[\ellref P]\dd x+\ln\frac{L}{\ellref}
=\ln\frac{L}{\ellref}-S,
\label{eq:KLexpand}
\end{equation}
so $S\le\ln(L/\ellref)$, with equality only when $P=P_u$ almost everywhere.

\section{First integrals of the regularized efficiency equations}
\label{app:regEta}

On the first half-cell, Eq.~\eqref{eq:regEta1} reads $2\mu y''+e^{-y}/B=0$. Multiplying by $y'$,
\begin{equation}
\mu\frac{\dd}{\dd x}\bigl(y'^2\bigr)
-\frac{1}{B}\frac{\dd}{\dd x}e^{-y}=0,
\label{eq:appFI1}
\end{equation}
so
\begin{equation}
\mu\,y'^2-\frac{e^{-y}}{B}=C_1 .
\label{eq:appFI2}
\end{equation}
On the second half-cell, Eq.~\eqref{eq:regEta2} gives, by the same manipulation,
\begin{equation}
\mu\,y'^2+\frac{A}{B^2}\,e^{-y}=C_2 .
\label{eq:appFI3}
\end{equation}
For fixed endpoints, $y$ is continuous, and the natural corner condition of the variational problem requires continuity of $y'$ at $x=L/2$. The constants $C_1$ and $C_2$, together with the self-consistent definitions of $A$ and $B$ on the solution, determine the smooth profile by a shooting method. In the small-$\mu$ limit the first integrals resolve the narrow transition layer near $L/2$; in the large-$\mu$ limit both reduce to nearly constant $y'$, i.e., the exponential profile.

\section{Numerical discretization and checks}
\label{app:numerics}

Divide the cell into $N$ elements of width $h=L/N$. On element $i$, take $T_i$ and $U_i'$ constant and define
\begin{equation}
g_i=\frac{U_i'}{T_i},
\qquad
\Phi_{i+1}=\Phi_i+g_ih .
\label{eq:appDisc1}
\end{equation}
The integrating-factor integral is evaluated exactly on each element,
\begin{equation}
\int_0^he^{\Phi_i+g_is}\dd s
=e^{\Phi_i}\,\frac{e^{g_ih}-1}{g_i},
\label{eq:appDisc2}
\end{equation}
with the limit $he^{\Phi_i}$ as $g_i\to0$. These exact element integrals construct the kernel $\cG(x)$ and the resistance $\cR$ of Eqs.~\eqref{eq:Gkernel}--\eqref{eq:R}; the midpoint density from Eq.~\eqref{eq:Pexact} is then used for $S$.

For the regularized fixed-endpoint optimization of Fig.~\ref{fig:reg}, the control is discretized on the nodal logarithmic variables $y_i=\ln T_i$ with $y_0=\ln\Th$ and $y_N=\ln\Tc$ held fixed and the interior nodes bounded by $\ln\Tc\le y_i\le\ln\Th$. The discrete objective $\eta-\mu\sum_i(y_{i+1}-y_i)^2/h$ is maximized by a bound-constrained quasi-Newton method, using the analytic efficiency gradient obtained by discretizing Eq.~\eqref{eq:gradEta} and applying the chain rule $\partial T/\partial y=T$, together with the standard quadratic-penalty gradient. An equivalent monotone parameterization writes the temperature drops as nonnegative weights $w_j$ with $\sum_jw_j=1$,
\begin{equation}
T_i=\Tc+(\Th-\Tc)
\Bigl(1-\sum_{j<i}w_j\Bigr),
\label{eq:appDisc3}
\end{equation}
which enforces $T_0=\Th$, $T_{N-1}=\Tc$, and $T_{i+1}\le T_i$ exactly. Grid refinement verifies convergence of $J$, $S$, and the roots of Eqs.~\eqref{eq:UoptExact} and \eqref{eq:UoptEP}.

The functional gradients were checked by centered finite differences. For a perturbation localized on element $i$, the difference quotient
\begin{equation}
\frac{J(T_i+\epsilon)-J(T_i-\epsilon)}{2\epsilon}
\label{eq:appFD}
\end{equation}
converges to Eq.~\eqref{eq:gradJ}, and the analogous entropy quotient converges to Eq.~\eqref{eq:gradS}. Constrained entropy optimization for $U_0>\Th-\Tc$ converges to the clipped form \eqref{eq:Tclip}, with the compensating slope on nonsaturated intervals.

\section{Stochastic convention and interface matching}
\label{app:convention}

A convenient way to expose which results are convention independent is the one-parameter family of currents
\begin{equation}
J_a=-\frac1\gamma\bigl[U'P+TP'+aT'P\bigr],
\qquad 0\le a\le1 .
\label{eq:JaFamily}
\end{equation}
The choices $a=1$, $a=1/2$, and $a=0$ correspond, respectively, to the divergence-form current used in Eq.~\eqref{eq:Jdef}, a midpoint thermal drift, and a Fick-type current. Define
\begin{equation}
Q_a=T^aP,
\qquad
\Phi(x)=\int_0^x\frac{U'(z)}{T(z)}\dd z .
\label{eq:Qa}
\end{equation}
The stationary equation becomes
\begin{equation}
Q_a'+\frac{U'}T\,Q_a=-\gamma J_a\,T^{a-1},
\label{eq:QaODE}
\end{equation}
so the quantity continuous across an ideal temperature jump is $T^aP$. Integration by the same integrating-factor construction as in Sec.~\ref{sec:current} gives
\begin{equation}
P_s^{(a)}(x)=\frac{e^{-\Phi(x)}\,\cG_a(x)}{T(x)^a\,\cR_a},
\qquad
J_a=\frac{1-e^{\Phi_L}}{\gamma\,\cR_a},
\label{eq:JaExact}
\end{equation}
where
\begin{align}
\cG_a(x)={}&\int_x^LT(y)^{a-1}e^{\Phi(y)}\dd y \nonumber\\
&+e^{\Phi_L}\int_0^xT(y)^{a-1}e^{\Phi(y)}\dd y,
\label{eq:Ga}\\
\cR_a={}&\int_0^L\frac{e^{-\Phi(x)}}{T(x)^a}\,\cG_a(x)\dd x .
\label{eq:Ra}
\end{align}
The sign and stall condition are independent of $a$,
\begin{equation}
\sgn J_a=-\sgn\Phi_L,
\qquad
J_a=0\;\Longleftrightarrow\;\oint\frac{U'}T\dd x=0,
\label{eq:aInvariant}
\end{equation}
so the stall force and the quasistatic efficiency functional of Secs.~\ref{sec:stall} and \ref{sec:etaopt} are convention independent within this family.

The finite-current results are not convention independent. Writing
\begin{equation}
q_a=\frac{T^aP_s}{J_a},
\qquad \nu=\frac1T,
\label{eq:qa}
\end{equation}
the state equation and normalization become
\begin{equation}
q_a'=-\gamma\,\nu^{1-a}-U'\nu\,q_a,
\qquad
J_a^{-1}=\int_0^L\nu^a q_a\dd x .
\label{eq:qaODE}
\end{equation}
The denominator, and therefore the response of the resistance to changes in $T(x)$, depends explicitly on $a$. Finite-current optimality conditions and profile rankings must consequently be evaluated with one fixed convention and its corresponding interface rule.

The uniform-density condition also changes with convention. Substituting $P_s=1/L$ into Eq.~\eqref{eq:JaFamily} gives
\begin{equation}
U_s'(x)+a\,T'(x)=0,
\label{eq:aUniform}
\end{equation}
so for $a>0$ the absolute Shannon-entropy maximizer is
\begin{equation}
T_{S,a}^*(x)=T_0-\frac{U_s(x)}a
\label{eq:TSa}
\end{equation}
whenever it obeys the bounds; Eq.~\eqref{eq:TSstar} is the $a=1$ member of this family.


\begin{thebibliography}{99}

\bibitem{Reimann2002}
P. Reimann, Brownian motors: Noisy transport far from equilibrium, Phys. Rep. \textbf{361}, 57 (2002).

\bibitem{Hanggi2009}
P. H\"anggi and F. Marchesoni, Artificial Brownian motors: Controlling transport on the nanoscale, Rev. Mod. Phys. \textbf{81}, 387 (2009).

\bibitem{Sekimoto2010}
K. Sekimoto, \textit{Stochastic Energetics} (Springer, Berlin, 2010).

\bibitem{Seifert2012}
U. Seifert, Stochastic thermodynamics, fluctuation theorems and molecular machines, Rep. Prog. Phys. \textbf{75}, 126001 (2012).

\bibitem{Landauer1975}
R. Landauer, Inadequacy of entropy and entropy derivatives in characterizing the steady state, Phys. Rev. A \textbf{12}, 636 (1975).

\bibitem{Buttiker1987}
M. B\"uttiker, Transport as a consequence of state-dependent diffusion, Z. Phys. B \textbf{68}, 161 (1987).

\bibitem{Derenyi1999}
I. Der\'enyi and R. D. Astumian, Efficiency of Brownian heat engines, Phys. Rev. E \textbf{59}, R6219 (1999).

\bibitem{Hondou2000}
T. Hondou and K. Sekimoto, Unattainability of Carnot efficiency in the Brownian heat engine, Phys. Rev. E \textbf{62}, 6021 (2000).

\bibitem{AsfawBekele2004}
M. Asfaw and M. Bekele, Current, maximum power and optimized efficiency of a Brownian heat engine, Eur. Phys. J. B \textbf{38}, 457 (2004).

\bibitem{Taye2017}
M. A. Taye, Irreversible Brownian heat engine, J. Stat. Phys. \textbf{169}, 423 (2017).

\bibitem{TayeExp2025}
M. A. Taye, Curzon--Ahlborn-type efficiency in a Brownian heat engine with exponential temperature profile, Phys. Rev. E \textbf{112}, 044122 (2025).

\bibitem{CurzonAhlborn1975}
F. L. Curzon and B. Ahlborn, Efficiency of a Carnot engine at maximum power output, Am. J. Phys. \textbf{43}, 22 (1975).

\bibitem{Berger2009}
F. Berger, T. Schmiedl, and U. Seifert, Optimal potentials for temperature ratchets, Phys. Rev. E \textbf{79}, 031118 (2009).

\bibitem{Polettini2013}
M. Polettini, Diffusion in nonuniform temperature and its geometric analog, Phys. Rev. E \textbf{87}, 032126 (2013).

\bibitem{Celani2012}
A. Celani, S. Bo, R. Eichhorn, and E. Aurell, Anomalous thermodynamics at the microscale, Phys. Rev. Lett. \textbf{109}, 260603 (2012).

\bibitem{BoCelani2013}
S. Bo and A. Celani, Entropic anomaly and maximal efficiency of microscopic heat engines, Phys. Rev. E \textbf{87}, 050102(R) (2013).

\bibitem{Taye2016}
M. A. Taye, Free energy and entropy production rate for a Brownian particle that walks on overdamped medium, Phys. Rev. E \textbf{94}, 032111 (2016).

\bibitem{Taye2020}
M. A. Taye, Entropy production and entropy extraction rates for a Brownian particle that walks in underdamped medium, Phys. Rev. E \textbf{101}, 012131 (2020).

\bibitem{Taye2021}
M. A. Taye, Effect of viscous friction on entropy, entropy production, and entropy extraction rates in underdamped and overdamped media, Phys. Rev. E \textbf{103}, 042132 (2021).

\bibitem{Taye2025CM}
M. A. Taye, Stochastic modeling of HIV reactivation under ART washout and immune fluctuations, Contemp. Math. \textbf{6}(5), 5708 (2025), doi:10.37256/cm.6520256801.

\bibitem{Perelson1993}
A. S. Perelson, D. E. Kirschner, and R. De Boer, Dynamics of HIV infection of CD4+ T cells, Math. Biosci. \textbf{114}, 81 (1993).

\bibitem{Perelson1996}
A. S. Perelson, A. U. Neumann, M. Markowitz, J. M. Leonard, and D. D. Ho, HIV-1 dynamics in vivo: virion clearance rate, infected cell life-span, and viral generation time, Science \textbf{271}, 1582 (1996).

\bibitem{Ho1995}
D. D. Ho, A. U. Neumann, A. S. Perelson, W. Chen, J. M. Leonard, and M. Markowitz, Rapid turnover of plasma virions and CD4 lymphocytes in HIV-1 infection, Nature (London) \textbf{373}, 123 (1995).

\bibitem{Wei1995}
X. Wei, S. K. Ghosh, M. E. Taylor, V. A. Johnson, E. A. Emini, P. Deutsch \textit{et al.}, Viral dynamics in human immunodeficiency virus type 1 infection, Nature (London) \textbf{373}, 117 (1995).

\bibitem{Chun1997}
T. W. Chun, L. Stuyver, S. B. Mizell, L. A. Ehler, J. A. Mican, M. Baseler \textit{et al.}, Presence of an inducible HIV-1 latent reservoir during highly active antiretroviral therapy, Proc. Natl. Acad. Sci. USA \textbf{94}, 13193 (1997).

\bibitem{Siliciano2003}
J. D. Siliciano, J. Kajdas, D. Finzi, T. C. Quinn, K. Chadwick, J. B. Margolick \textit{et al.}, Long-term follow-up studies confirm the stability of the latent reservoir for HIV-1 in resting CD4+ T cells, Nat. Med. \textbf{9}, 727 (2003).

\bibitem{Murray2016}
A. J. Murray, K. J. Kwon, D. L. Farber, and R. F. Siliciano, The latent reservoir for HIV-1: how immunologic memory and clonal expansion contribute to HIV-1 persistence, J. Immunol. \textbf{197}, 407 (2016).

\bibitem{Li2020}
J. Z. Li \textit{et al.}, Proviruses with identical sequences comprise a large fraction of the replication-competent HIV reservoir, Proc. Natl. Acad. Sci. USA \textbf{117}, 3886 (2020).

\bibitem{Hill2014}
A. L. Hill, D. I. S. Rosenbloom, F. Fu, M. A. Nowak, and R. F. Siliciano, Predicting the outcomes of treatment to eradicate the latent reservoir for HIV-1, Proc. Natl. Acad. Sci. USA \textbf{111}, 13475 (2014).

\bibitem{Hill2018}
A. L. Hill, D. I. S. Rosenbloom, E. Goldberg, E. Hanhauser, D. R. Kuritzkes, R. F. Siliciano \textit{et al.}, Real-time predictions of reservoir size and rebound time during antiretroviral therapy interruption trials for HIV, PLoS Pathog. \textbf{14}, e1007333 (2018).

\bibitem{pinkevych2015latency}
M. Pinkevych, D. Cromer, M. Tolstrup, A. J. Grimm, D. A. Cooper, S. R. Lewin, O. S. S{\o}gaard, T. A. Rasmussen, S. J. Kent, A. D. Kelleher, and M. P. Davenport, HIV reactivation from latency after treatment interruption occurs on average every 5--8 days: implications for HIV remission, PLoS Pathog. \textbf{11}, e1005000 (2015).

\bibitem{conway2019rebound}
J. M. Conway, A. S. Perelson, and J. Z. Li, Predictions of time to HIV viral rebound following ART suspension that incorporate personal biomarkers, PLoS Comput. Biol. \textbf{15}, e1007229 (2019).

\bibitem{NowakBangham1996}
M. A. Nowak and C. R. M. Bangham, Population dynamics of immune responses to persistent viruses, Science \textbf{272}, 74 (1996).

\bibitem{Perelson2002}
A. S. Perelson, Modelling viral and immune system dynamics, Nat. Rev. Immunol. \textbf{2}, 28 (2002).

\bibitem{Wodarz2002}
D. Wodarz and M. A. Nowak, Mathematical models of HIV pathogenesis and treatment, BioEssays \textbf{24}, 1178 (2002).

\bibitem{gunst2025ati}
J. D. Gunst, J. Gohil, J. Z. Li, R. J. Bosch \textit{et al.}, Time to HIV viral rebound and frequency of post-treatment control after analytical interruption of antiretroviral therapy: an individual data-based meta-analysis of 24 prospective studies, Nat. Commun. \textbf{16}, 906 (2025).

\bibitem{fennessey2017barcoded}
C. M. Fennessey, M. Pinkevych, T. T. Immonen, A. Reynaldi, V. Venturi, P. Nadella, C. Reid, L. Newman, L. Lipkey, K. Oswald \textit{et al.}, Genetically-barcoded SIV facilitates enumeration of rebound variants and estimation of reactivation rates in nonhuman primates following interruption of suppressive antiretroviral therapy, PLoS Pathog. \textbf{13}, e1006359 (2017).

\bibitem{VanDorp2020}
C. H. Van Dorp, J. M. Conway, D. H. Barouch, J. B. Whitney, and A. S. Perelson, Models of SIV rebound after treatment interruption that involve multiple reactivation events, PLoS Comput. Biol. \textbf{16}, e1008241 (2020).

\bibitem{li2016reservoir}
J. Z. Li, B. Etemad, H. Ahmed, E. Aga, R. J. Bosch, J. W. Mellors, D. R. Kuritzkes, M. M. Lederman, M. Para, and R. T. Gandhi, The size of the expressed HIV reservoir predicts timing of viral rebound after treatment interruption, AIDS \textbf{30}, 343 (2016).

\bibitem{pasternak2020carna}
A. O. Pasternak, S. DeMaster \textit{et al.}, Cell-associated HIV-1 RNA predicts viral rebound and disease progression after discontinuation of temporary early ART, JCI Insight \textbf{5}, e134196 (2020).

\bibitem{sneller2020kinetics}
M. C. Sneller, E. D. Huiting, K. E. Clarridge, C. Seamon, J. Blazkova, J. S. Justement \textit{et al.}, Kinetics of plasma HIV rebound in the era of modern antiretroviral therapy, J. Infect. Dis. \textbf{222}, 1655 (2020).

\bibitem{Pinkevych2019}
M. Pinkevych, D. Cromer, M. P. Davenport \textit{et al.}, Modeling of experimental data supports HIV reactivation from latency after treatment interruption at high but variable rates, eLife \textbf{8}, e49022 (2019).

\bibitem{Wu2020}
Y. Wu, M. Pinkevych, Z. Xu, B. F. Keele, M. P. Davenport, and D. Cromer, Impact of fluctuation in frequency of human immunodeficiency virus/simian immunodeficiency virus reactivation during antiretroviral therapy interruption, Proc. R. Soc. B \textbf{287}, 20200354 (2020).

\bibitem{Kingman1993}
J. F. C. Kingman, \textit{Poisson Processes} (Oxford University Press, Oxford, 1993).

\bibitem{Feller1971}
W. Feller, \textit{An Introduction to Probability Theory and Its Applications}, Vol. II, 2nd ed. (Wiley, New York, 1971).

\bibitem{CoxMiller1965}
D. R. Cox and H. D. Miller, \textit{The Theory of Stochastic Processes} (Chapman and Hall, London, 1965).

\bibitem{Taye2026StochasticFirstPassage}
M. A. Taye, Stochastic first-passage theory of HIV viral rebound following latent reservoir reactivation, arXiv:2607.04910 (2026).

\bibitem{Taye2026BiologicalTimeEvolutionary}
M. A. Taye, Biological time, evolutionary optimization, and gauge coherence: A thermodynamic synthesis of the principle of biological time equivalence, arXiv:2607.04827 (2026).

\bibitem{Taye2026RelativisticPbteBiological}
M. A. Taye, Relativistic PBTE: Biological proper time along the worldline, arXiv:2607.04849 (2026).

\bibitem{Taye2026NonequilibriumInternalTime}
M. Taye, A nonequilibrium internal-time model of aging: Entropy-normalized biological proper time and repair bifurcations, arXiv:2606.23279 (2026).

\bibitem{Taye2026BiologicalProperTime}
M. A. Taye, Int. J. Sci. Res. Publ. \textbf{16}, 17417 (2026).

\bibitem{Taye2026PrincipleBiologicalTime}
M. A. Taye, The principle of biological time equivalence (2026).

\bibitem{Taye2026NonequilibriumThermodynamicsStochastic}
M. A. Taye, Nonequilibrium thermodynamics in stochastic processes (2026).

\bibitem{Taye2026BrownianMotorsBrownian}
M. A. Taye, \textit{Brownian Motors and Brownian Heat Engines: From Classical Thermodynamics to Fluctuation-Driven Machines} (Independently published, 2026).

\bibitem{Taye2026ExactThermodynamicAnalysis}
M. A. Taye, Exact thermodynamic analysis of a hybrid molecular motor switching between active and passive modes (2026).

\bibitem{Taye2026NoiseActivatedDopant}
M. A. Taye, Noise-activated dopant dynamics in two-dimensional thermal landscapes with localized cold spots, Int. J. Sci. Res. Publ. \textbf{16}(5) (2026).

\bibitem{Taye2026UniversalThermodynamicInequality}
M. A. Taye, A universal thermodynamic inequality: Scaling relations between current, activity, and entropy production (2026).

\bibitem{Taye2026ThermodynamicParametrisationVertebrate}
M. A. Taye, Int. J. Sci. Res. Publ. \textbf{16}, 2250 (2026).

\bibitem{Taye2026NeuralInvestmentEntropy}
M. Taye, Neural investment as an entropy-budget strategy: A thermodynamic derivation of primate longevity from the principle of biological time equivalence, arXiv:2604.27937 (2026).

\bibitem{Taye2026LifetimeCardiacCycle}
M. Taye, The lifetime cardiac-cycle invariant in endothermic vertebrates: A 230-species comparative dataset, statistical validation, and explicit falsifiability criteria, arXiv:2604.27856 (2026).

\bibitem{Taye2026BiologicalTimeEquivalence}
M. Taye, Biological time equivalence in vertebrates: Thermodynamic framework, comparative tests, and clade-specific deviations, arXiv:2603.26377 (2026).

\bibitem{Taye2026EntropyProductionMacroscopic}
M. A. Taye, Mod. Math. Phys. \textbf{2}, 1 (2026).

\bibitem{Taye2025CompetingActivePassive}
M. A. Taye, Physica A, 131214 (2025).

\bibitem{Taye2025ThermodynamicIrreversibilityUnderdamped}
M. A. Taye, Phys. Rev. E \textbf{112}, 044101 (2025).

\bibitem{Taye2025UnifiedNonequilibriumFramework}
M. Taye, A unified nonequilibrium framework: Thermodynamic distance, dissipation, and stationary laws via effective state count, variational stationarity, and thermodynamic bounds, arXiv:2509.09041 (2025).

\bibitem{taye2025EntropyProductionThermodynamic}
M. A. Taye, Contemp. Math. \textbf{6}, 4101 (2025).

\bibitem{Taye2025ThermodynamicFeaturesHeat}
M. Taye, Thermodynamic features of a heat engine coupled with exponentially decreasing temperature across the reaction coordinate, as well as perspectives on nonequilibrium thermodynamics, arXiv:2503.24317 (2025).

\bibitem{Taye2025DrugWashoutViral}
M. A. Taye, bioRxiv 2025.03.22.644757 (2025).

\bibitem{Taye2025ThermodynamicRelationsTime}
M. Taye, arXiv:2503.20812 (2025).

\bibitem{Taye2025EntropyProductionThermodynamic}
M. A. Taye, Entropy production and thermodynamic dynamics in active and passive Brownian systems driven by time-dependent forces and temperatures (2025).

\bibitem{Taye2025DirectedTransportShort}
M. A. Taye, Directed transport of a short polymer chain on a temperature-dependent ratchet potential (2025).

\bibitem{Taye2024ExactTimeDependent}
M. A. Taye, Phys. Rev. E \textbf{110}, 054105 (2024).

\bibitem{Taye2024ExactTimeDependent2}
M. A. Taye, Contemp. Math., 5113 (2024).

\bibitem{Taye2023TimeDependentThermodynamic}
M. A. Taye, bioRxiv 2023.12.06.570486 (2023).

\bibitem{Mahmud2023ComputationalInvestigationCis}
M. Mahmud, M. Bekele, and N. Behera, Theory Biosci. \textbf{142}, 151 (2023).

\bibitem{Ashagre2023SynergisticContributionPotassium}
S. Ashagre, A. K. Ogundele, J. N. Ike, B. Gebremichael, M. Bekele, G. D. Sharma \textit{et al.}, J. Phys. Chem. Solids \textbf{177}, 111290 (2023).

\bibitem{Taye2023TimeDependentSolutions}
M. A. Taye, Eur. Phys. J. B \textbf{96}, 65 (2023).

\bibitem{Taye2023TransportingShortPolymer}
M. Taye, Transporting a short polymer along a reaction coordinate that coupled with a spatially varying temperature (2023).

\bibitem{Taye2023DynamicsBloodCells}
M. A. Taye, bioRxiv 2023.01.21.525013 (2023).

\bibitem{Taye2023HostViralLoad}
M. A. Taye, Contemp. Math. \textbf{4}, 392 (2023).

\bibitem{Mahmud2022ThermalQuantumFluctuation}
M. Mahmud, M. Bekele, and Y. Bassie, Condens. Matter \textbf{7}, 62 (2022).

\bibitem{Taye2022ExactTimeDependent2}
M. A. Taye, Phys. Rev. E \textbf{105}, 054126 (2022).

\bibitem{Aragie2022NoiseFormedTriple}
B. Aragie, M. Bekele, and G. Pellicane, Pramana \textbf{96}, 59 (2022).

\bibitem{Abebe2022ThermallyActivatedDiffusion}
Y. Abebe, T. Birhanu, L. Demeyu, M. Taye, M. Bekele, and Y. Bassie, Eur. Phys. J. B \textbf{95}, 9 (2022).

\bibitem{Birhanu2021StochasticResonatorLayered}
T. Birhanu, Y. Abebe, L. Demeyu, M. Taye, and M. Bekele, Int. J. Mod. Phys. B \textbf{35}, 2150284 (2021).

\bibitem{Taye2021BrownianMotorsArranged}
M. A. Taye, Eur. Phys. J. B \textbf{94}, 124 (2021).

\bibitem{Zhang2021EfficacyAntiviralDrug}
S. Y. Zhang and M. A. Taye, The efficacy of antiviral drug, HIV viral load and the immune response, arXiv:2101.10413 (2021).

\bibitem{Taye2020CorrelationBetweenAntiviral}
M. A. Taye, bioRxiv 2020.11.06.372094 (2020).

\bibitem{Taye2020SedimentationRateErythrocyte}
M. A. Taye, Eur. Phys. J. E \textbf{43}, 19 (2020).

\bibitem{Taye2019PhysicsErythrocyteSedimentation}
M. A. Taye, The physics of erythrocyte sedimentation rate, arXiv:1907.12148 (2019).

\bibitem{Duki2018StochasticResonanceFirst2}
S. F. Duki and M. A. Taye, J. Stat. Phys. \textbf{171}, 878 (2018).

\bibitem{Duki2016FirstPassageTime}
S. F. Duki and M. A. Taye, First passage time and stochastic resonance of excitable systems, arXiv:1609.07752 (2016).

\bibitem{Taye2015EffectTemperatureDependence}
M. A. Taye and S. F. Duki, Eur. Phys. J. B \textbf{88}, 322 (2015).

\bibitem{Taye2015ExactAnalyticalThermodynamic}
M. A. Taye, Phys. Rev. E \textbf{92}, 032126 (2015).

\bibitem{Taye2015RectifiedMotionShort}
M. A. Taye, Rectified motion of short polymer chain that walks along a ratchet potential that coupled with spatially varying temperature, arXiv:1507.04945 (2015).

\bibitem{Duki2015EffectTemperatureViscous}
S. F. Duki and M. A. Taye, The effect of temperature on viscous friction and the performance of a Brownian heat engine, arXiv:1507.02005 (2015).

\bibitem{Taye2015ExactAnalyticalExpressions}
M. A. Taye, Exact analytical expressions for entropy production and free energy, arXiv:1507.01791 (2015).

\bibitem{Aragie2014ImpurityDiffusionHarmonic}
B. Aragie, M. Asfaw, L. Demeyu, and M. Bekele, Eur. Phys. J. B \textbf{87}, 214 (2014).

\bibitem{Asfaw2014ThermallyActivatedBarrier}
M. Asfaw, Thermally activated barrier crossing rate for a coupled system moving in a ratchet potential, arXiv:1407.3713 (2014).

\bibitem{Asfaw2014ThermodynamicFeatureBrownian}
M. Asfaw, Phys. Rev. E \textbf{89}, 012143 (2014).

\bibitem{Asfaw2013CorrectionTimingStatistics}
M. Asfaw, E. Alvarez-Lacalle, and Y. Shiferaw, PLoS ONE \textbf{8} (2013), doi:10.1371/annotation/10d4ef64-c7e6-43ff-8bd7-658d47689855.

\bibitem{Asfaw2013TimingStatisticsSpontaneous}
M. Asfaw, E. Alvarez-Lacalle, and Y. Shiferaw, PLoS ONE \textbf{8}, e62967 (2013).

\bibitem{Asfaw2013EffectThermalInhomogeneity}
M. Asfaw, Eur. Phys. J. B \textbf{86}, 189 (2013).

\bibitem{Chen2012StatisticsCalciumMediated}
W. Chen, M. Asfaw, and Y. Shiferaw, Biophys. J. \textbf{102}, 461 (2012).

\bibitem{Shiferaw2012CalciumMediatedTriggered}
Y. Shiferaw, W. Chen, and M. Asfaw, Biophys. J. \textbf{102}, 672a (2012).

\bibitem{Asfaw2012ExploringDynamicsDimer}
M. Asfaw and Y. Shiferaw, Exploring the dynamics of dimer crossing over a Kramers type potential, J. Chem. Phys. \textbf{136}, 025101 (2012).

\bibitem{Asfaw2011ExploringElasticFeatures}
M. Asfaw, J. Phys.: Condens. Matter \textbf{23}, 105101 (2011).

\bibitem{Asfaw2011NoiseCreatedBistability}
M. Asfaw, B. Aragie, and M. Bekele, Eur. Phys. J. B \textbf{79}, 371 (2011).

\bibitem{Asfaw2011LateralPhaseSeparation}
M. Asfaw, Int. J. Mod. Phys. B \textbf{25}, 457 (2011).

\bibitem{Asfaw2010ThermallyActivatedBarrier}
M. Asfaw, Phys. Rev. E \textbf{82}, 021111 (2010).

\bibitem{Asfaw2010StochasticResonanceFlexible}
M. Asfaw and W. Sung, EPL \textbf{90}, 30008 (2010).

\bibitem{Bekele2009EvaluationClimateChange}
H. M. Bekele, Evaluation of climate change impact on upper Blue Nile Basin reservoirs, M.S. thesis, Arba Minch University (2009).

\bibitem{Asfaw2008AdhesionInducedLateral}
M. Asfaw and H. Y. Chen, Adhesion-induced lateral phase separation of multi-component membranes: the effect of repellers and confinement, arXiv:0811.3725 (2008).

\bibitem{Asfaw2008ModelingEfficientBrownian}
M. Asfaw, Eur. Phys. J. B \textbf{65}, 109 (2008).

\bibitem{Getahun2008CompetingJumpCycles}
Z. Getahun, M. Asfaw, and M. Bekele, Competing jump cycles for vacancy diffusion in binary alloys, arXiv:0807.5034 (2008).

\bibitem{Asfaw2008UnbindingTransitionsMembranes}
M. Asfaw, Physica A \textbf{387}, 3526 (2008).

\bibitem{Asfaw2007AdhesionInducedLateral}
M. Asfaw, Bussei Kenky\={u} \textbf{89}, 12 (2007).

\bibitem{Demeyu2007MonteCarloSimulations}
L. Demeyu, S. Stafstr\"om, and M. Bekele, Phys. Rev. B \textbf{76}, 155202 (2007).

\bibitem{Asfaw2007ExploringOperationTiny}
M. Asfaw and M. Bekele, Physica A \textbf{384}, 346 (2007).

\bibitem{Asfaw2007UnbindingTransitionsMulticomponent}
M. Asfaw, Unbinding transitions of multicomponent membranes and strings, arXiv:0706.3433 (2007).

\bibitem{Asfaw2006MembraneAdhesionVia}
M. Asfaw, B. R\'o\.zycki, R. Lipowsky, and T. R. Weikl, EPL \textbf{76}, 703 (2006).

\bibitem{Asfaw2005EnergeticsSimpleMicroscopic}
M. Asfaw and M. Bekele, Phys. Rev. E \textbf{72}, 056109 (2005).

\bibitem{Asfaw2005AdhesionMultiComponent}
M. Asfaw, Adhesion of multi-component membranes and strings (2005).

\bibitem{Pearson2003ManagementHealthReproduction}
R. A. Pearson, M. Alemayehu, A. Tesfaye, D. G. Smith, G. Kebede, and M. Asfaw, Management, health and reproduction of donkeys used for work in peri-urban areas of West and East Shewa, Ethiopia: a survey (2003).

\bibitem{PEARSON2003UseManagementDonkeys}
R. A. Pearson, M. Alemayehu, A. Tesfaye, E. F. Allan, D. G. Smith \textit{et al.}, Use and management of donkeys in peri-urban areas of Ethiopia, Phase One (2003).

\bibitem{Asfaw2002AdjustableBrownianHeat}
M. Asfaw and M. Bekele, An adjustable Brownian heat engine, arXiv:cond-mat (2002).

\end{thebibliography}
\end{document}